\documentclass[fleqn,usenatbib]{mnras}

\usepackage{newtxtext,newtxmath}

\usepackage[T1]{fontenc}

\DeclareRobustCommand{\VAN}[3]{#2}
\let\VANthebibliography\thebibliography
\def\thebibliography{\DeclareRobustCommand{\VAN}[3]{##3}\VANthebibliography}

\usepackage{graphicx}	
\usepackage{amsmath}	
\usepackage{amssymb}	
\usepackage{natbib}
\usepackage{url}
\usepackage{longtable}
\usepackage{threeparttable}
\usepackage{footnote}
\newcommand{\source}{MAXI~J1535$-$571}

\newcommand{\arcsecond}{$^{\prime\prime}$}

\newcommand{\swift}{\textit{Swift}}

\newcommand{\nh}{$N_{\mathrm{H}}$}
\newcommand{\nub}{$\nu_{\mathrm{break}}$}
\newcommand{\rz}{$R_{\mathrm{F}}$}
\newcommand{\bz}{$B_{\mathrm{F}}$}
\newcommand{\rg}{$r_{\mathrm{g}}$}
\graphicspath{{./}{Figures/}}

\title[Rapid compact jet quenching in MAXI J1535$-$571]{Rapid compact jet quenching in the Galactic black hole candidate X-ray binary MAXI~J1535$-$571}

\author[T. D. Russell et al.]{T. D. Russell,$^{1}$\thanks{E-mail: t.d.russell@uva.nl}
M. Lucchini,$^{1}$
A. J. Tetarenko,$^{2,3}$
J. C. A. Miller-Jones,$^{4}$
G. R. Sivakoff,$^{3}$
\newauthor F. Krau{\ss},$^{5,6}$
W. Mulaudzi,$^{7}$
M. C. Baglio,$^{8,9}$
D. M. Russell,$^{8}$
D. Altamirano,$^{10}$
C. Ceccobello,$^{11}$
\newauthor S. Corbel,$^{12,13}$
N. Degenaar,$^{1}$
J. van den Eijnden,$^{1}$
R. Fender,$^{14}$
S. Heinz,$^{15}$
K. I. I. Koljonen,$^{16,17}$
\newauthor 
D. Maitra,$^{18}$
S. Markoff,$^{1}$
S. Migliari,$^{19,20}$
A. S. Parikh,$^{1}$
R. M. Plotkin,$^{21}$
M. Rupen,$^{22}$
\newauthor C. Sarazin,$^{23}$
R. Soria,$^{24,25}$
and R. Wijnands$^{1}$
\\
$^{1}$Anton Pannekoek Institute for Astronomy, University of Amsterdam, Science Park 904, NL-1098 XH Amsterdam, The Netherlands\\
$^{2}$East Asian Observatory, 660 N. A'oh\={o}k\={u} Place, University
Park, Hilo, Hawaii 96720, USA\\
$^{3}$Department of Physics, University of Alberta, CCIS 4-181, Edmonton, AB T6G 2E1, Canada\\
$^{4}$International Centre for Radio Astronomy Research - Curtin University, GPO Box U1987, Perth, WA 6845, Australia\\
$^{5}$GRAPPA \& API, University of Amsterdam, Science Park 904, NL-1098 XH Amsterdam, The Netherlands\\
$^{6}$Department of Astronomy \& Astrophysics, Pennsylvania State University, University Park, PA, 16802, USA\\
$^{7}$Department of Astronomy, University of Cape Town, Private Bag X3, Rondebosch 7701, South Africa\\
$^{8}$Center for Astro, Particle and Planetary Physics, New York University Abu Dhabi, PO Box 129188, Abu Dhabi, UAE\\
$^{9}$INAF, Osservatorio Astronomico di Brera, Via E. Bianchi 46, 23807 Merate, LC, Italy\\ 
$^{10}$School of Physics and Astronomy, University of Southampton, Highfield SO17 IBJ, England\\
$^{11}$Department of Space, Earth and Environment, Chalmers University of Technology, Onsala Space Observatory, 439 92 Onsala, Sweden\\
$^{12}$AIM, CEA, CNRS, Universit\'{e} Paris Diderot, Sorbonne Paris Cit\'{e}, Universit\'{e} Paris-Saclay, F-91191 Gif-sur-Yvette, France\\
$^{13}$Station de Radioastronomie de Nan\c{c}ay, Observatoire de Paris, PSL Research University, CNRS, Univ. Orl\'eans, 18330 Nan\c{c}ay, France\\
$^{14}$Astrophysics, Department of Physics, Denys Wilkinson Building, Keble Road, Oxford OX1 3RH, UK\\
$^{15}$Department of Astronomy, University of Wisconsin-Madison, 475 N. Charter Street, Madison, WI 53706, USA\\
$^{16}$Finnish Centre for Astronomy with ESO (FINCA), Vesilinnantie 5, FI-20014 University of Turku, Finland\\
$^{17}$Aalto University Mets\"ahovi Radio Observatory, Mets\"ahovintie 114, FI-02540 Kylm\"al\"a, Finland\\
$^{18}$Department of Physics and Astronomy, Wheaton College, Norton, MA 02766, USA\\
$^{19}$XMM-Newton Science Operations Centre, ESAC/ESA, Camino Bajo del Castillo s/n, Urb. Villafranca del Castillo, 28691 Villanueva de la Ca\~nada, Madrid, Spain\\
$^{20}$Institute of Cosmos Sciences, University of Barcelona, Mart\'i Franqu\'es 1, 08028 Barcelona, Spain\\
$^{21}$Department of Physics, University of Nevada, Reno, Nevada 89557, USA\\
$^{22}$Herzberg Institute of Astrophysics, National Research Council of Canada, Penticton, BC, Canada\\
$^{23}$Department of Astronomy, University of Virginia, 530 McCormick Road, Charlottesville, VA 22904-4325, USA\\
$^{24}$College of Astronomy and Space Sciences, University of the Chinese Academy of Sciences, Beijing 100049, China\\
$^{25}$Sydney Institute for Astronomy, School of Physics A28, The University of Sydney, Sydney, NSW, 2006, Australia
}

\date{Accepted XXX. Received YYY; in original form ZZZ}

\pubyear{2020}

\begin{document}
\label{firstpage}
\pagerange{\pageref{firstpage}--\pageref{lastpage}}
\maketitle

\begin{abstract}
We present results from six epochs of quasi-simultaneous radio, (sub-)millimetre, infrared, optical, and X-ray observations of the black hole X-ray binary MAXI~J1535$-$571. These observations show that as the source transitioned through the hard-intermediate X-ray state towards the soft intermediate X-ray state, the jet underwent dramatic and rapid changes. We observed the frequency of the jet spectral break, which corresponds to the most compact region in the jet where particle acceleration begins (higher frequencies indicate closer to the black hole), evolve from the IR band into the radio band (decreasing by $\approx$3 orders of magnitude) in less than a day. During one observational epoch, we found evidence of the jet spectral break evolving in frequency through the radio band. Estimating the magnetic field and size of the particle acceleration region shows that the rapid fading of the high-energy jet emission was not consistent with radiative cooling; instead the particle acceleration region seems to be moving away from the black hole on approximately dynamical timescales. This result suggests that the compact jet quenching is not caused by local changes to the particle acceleration, rather we are observing the acceleration region of the jet travelling away from the black hole with the jet flow. Spectral analysis of the X-ray emission show a gradual softening in the few days before the dramatic jet changes, followed by a more rapid softening $\sim$1--2\,days after the onset of the jet quenching.

\end{abstract}

\begin{keywords}
X-rays: binaries -- accretion, accretion disks -- acceleration of particles -- ISM: jets and outflows -- submillimetre: general --  X-rays: individual (MAXI J1535$-$571)
\end{keywords}


\section{Introduction}

Accreting stellar-mass black holes (BHs) in X-ray binaries (XRBs) are able to launch powerful, collimated outflows, or jets, which dominate the observed emission at radio to infrared (IR) wavelengths. While there is an observable connection between the processes of accretion and jet production in BH XRBs \citep[e.g.,][]{2004ApJ...617.1272C,2004MNRAS.355.1105F}, at present, precisely how these jets are launched, and the detailed coupling with the accretion flow remain important yet poorly understood astrophysical questions.

During periods of increased mass accretion onto the BH, XRBs go through phases of bright outburst where they evolve through their distinct modes of accretion on timescales of weeks to months (or even years). During such outbursts the observed properties of the jets change dramatically \citep[e.g.,][]{2006csxs.book..381F}. Therefore, simultaneous multiwavelength observations of these systems that monitor both the jets and the accretion flow (observed at optical, X-ray and higher frequencies) during their outbursts provides a unique view of the accretion-jet evolution, probing the structural changes in the jet, and helping to constrain the underlying jet physics \citep[e.g.,][]{2000A&A...359..251C,2003A&A...400.1007C,2005ApJ...635.1203M,2011A&A...529A...3C,2014MNRAS.439.1390R,2018MNRAS.473.4417C}.

In an outburst, BH XRBs are initially in a hard X-ray spectral state (see \citealt{2010LNP...794...53B} for a review of the X-ray accretion states) where the X-ray emission is dominated by a power-law component from inverse Compton scattering by hot electrons in the innermost regions \citep[e.g.,][]{1995ApJ...452..710N}. As the accretion rate increases the source transitions towards a soft X-ray spectral state, first moving through the hard- and soft-intermediate states (HIMS and SIMS, respectively). This evolution occurs as the the X-ray spectrum becomes progressively more dominated by soft X-ray emission from the accretion disk, while the hard power-law component becomes steeper (softens), due to either the emission region contracting \citep{2019Natur.565..198K} or a change in the spectrum as the jet evolves \citep[e.g.,][]{2013ApJ...773...59P}. After a few weeks to months (or even years), the mass accretion rate drops and the outburst begins to fade. The XRB then transitions back through the intermediate states to the hard state as the disk cools and the power-law component begins to dominate once again.

Over the transition between the different accretion states, the observed jet properties change dramatically. The hard state is characterised by a steady, partially self-absorbed compact jet \citep[e.g.,][]{2000ApJ...543..373D,2000A&A...359..251C,2001MNRAS.322...31F,2001MNRAS.327.1273S,2004MNRAS.355.1105F}. Observed from radio to IR wavelengths, the compact jet exhibits a flat-to-inverted radio to mm spectrum ($\alpha \gtrsim 0$, where the observed flux density, $S_{\nu}$ scales with frequency, $\nu$, such that $S_{\nu} \propto \nu^{\alpha}$; e.g.\ \citealt{2001MNRAS.322...31F}). The flat spectrum extends up to $\sim$10$^{13}$\,Hz \citep[e.g.,][]{2002ApJ...573L..35C,2013MNRAS.429..815R}, above which the jet becomes optically thin and the spectrum is steep ($\alpha \approx -0.6$; e.g.,\ \citealt{2013MNRAS.429..815R}). This jet spectral break corresponds to the most compact region in the jet where particle acceleration begins (the first acceleration zone, e.g.,\ \citealt{2001A&A...372L..25M,2002ApJ...573L..35C,2005ApJ...635.1203M,2010LNP...794..143M,2017SSRv..207....5R}). The frequency (\nub) and flux density of the jet spectral break are likely set by internal jet plasma properties, and its frequency scales with its distance from the BH (with higher frequencies probing regions closer to the BH; \citealt{1979ApJ...232...34B}). Determining the break's location and evolution can help to reveal key properties of the jet outflow (e.g.\ \citealt{2003MNRAS.343L..59H,2011A&A...529A...3C,2014MNRAS.439.1390R,2014MNRAS.438..959P,2018MNRAS.473.4417C}).

In the early phase of a typical outburst, as the source transitions through the hard and then the intermediate states, the jet break is believed to gradually shift to lower frequencies, eventually moving through the radio band. One interpretation of this evolution is that the particle-accelerating region moves away from the BH\footnote{A change in the location of the jet spectral break has also been inferred in the neutron star XRB Aquila X-1 \citep{2018A&A...616A..23D}.}  (e.g.,\ \citealt{2011A&A...529A...3C,2014MNRAS.439.1390R}), due to changes in the jet internal properties that in turn dictate where particle acceleration happens (e.g.,\ \citealt{2014MNRAS.443..299M}). However, the detailed behaviour of the jet break during the rise has not yet been well determined, where the evolution of \nub\ during the outburst rise has only been inferred from observations of the radio spectrum changing from flat to steep as the jet break moves from above to below the radio band \citep[e.g.,][]{2013MNRAS.428.2500C}. \citet{2013MNRAS.436.2625V} presented radio monitoring of MAXI~J1659$-$152 showing that a single power law did not well represent the radio data. Their results suggested that the spectral break was moving from higher frequencies into the radio band and back to higher frequencies on $\sim$day timescales as the source transitioned back and forth between the HIMS, SIMS, and soft states.

At some point during the transition from the hard to soft state, the compact jet emission switches off \citep[e.g.,][]{2004MNRAS.355.1105F}, being quenched by at least 3.5 orders of magnitude \citep{2019ApJ...883..198R}. \nub\ evolving to lower frequencies signals the progressive quenching of the higher energy emission from the compact jet \citep{2012MNRAS.421..468M,2013MNRAS.428.2500C,2014MNRAS.439.1390R,2018A&A...614L...5K}. However, the complete evolution has not been directly observed. Around the transition to the soft state, a bright, transient jet can also be launched \citep[e.g.,][]{1994Natur.371...46M,1995Natur.375..464H,1995Natur.374..141T}, which is characterised by rapid flaring and an optically-thin radio spectrum \citep{1995Natur.375..464H,1999MNRAS.304..865F,2001MNRAS.322...31F}.  While no compact jet is observed in the soft state, residual radio emission may be observed from ejecta launched from the system around the state transition (transient jet; e.g., \citealt{2002Sci...298..196C,2004ApJ...617.1272C,2004MNRAS.355.1105F,2019ApJ...883..198R,2020NatAs.tmp....2B}). Additionally, from IR monitoring of the BH XRB 4U 1543$-$47, \citet{2020arXiv200208399R} reported an IR flare as the source briefly returned to the SIMS from the soft state. This flare suggested that the compact jet emission switched back on during this brief ($\sim$5\,day) return to the SIMS, before the source transitioned back to the soft state.

Towards the end of an outburst, the source moves back towards the hard state and the compact jet gradually re-establishes over a period of several weeks \citep{2013ApJ...779...95K,2014MNRAS.439.1390R}. The jet first brightens at lower-frequencies before brightening at IR and optical frequencies \citep{2012MNRAS.421..468M,2013MNRAS.431L.107C}. This progressive brightening is associated with X-ray spectral hardening, where \nub\ has been observed to shift to higher frequencies, first through the radio band (as shown by the radio spectrum evolving from steep to flat/inverted; e.g.,\ \citealt{2013MNRAS.431L.107C}), and then gradually up to IR frequencies \citep{2013ApJ...768L..35R,2014MNRAS.439.1390R}. The best sampled evolution of \nub\ to date was observed during this phase of an outburst from MAXI~J1836$-$194, where the break was observed to shift gradually by $\sim$3 orders of magnitude from low to higher frequencies over $\sim$6\,weeks, possibly connected to changes in the source hardness  \citep[][]{2014MNRAS.439.1390R}. 

\source\ is a Galactic black hole candidate X-ray binary that was first discovered in 2017 September, in the early phase of its $\sim$1-year long outburst \citep{2017ATel10699....1N,2017GCN.21788....1M}. This outburst was extensively monitored at radio \citep{2019ApJ...883..198R,2019ApJ...878L..28P,2019MNRAS.488L.129C}, sub-mm \citep{2017ATel10745....1T}, IR and optical \citep{2018ApJ...867..114B}, and X-ray \citep{2018MNRAS.480.4443T,2018PASJ...70...95N,2018ApJ...866..122H,2018ApJ...865L..15S,2019MNRAS.487..928S,2019MNRAS.488..720B,2019MNRAS.487.4221S} wavelengths. See \citet{2018MNRAS.480.4443T} and \citet{2018PASJ...70...95N} for a full discussion on the X-ray states during the outburst. \source\ is located 4.1$^{+0.6}_{-0.5}$\,kpc away (determined via an H\,I absorption study; \citealt{2019MNRAS.488L.129C}).

In this work, we present six epochs of quasi-simultaneous (typically $\pm$0.5\,day)\footnote{While this was adhered to at times of rapid changes, at times where the jet and accretion properties were not evolving as rapidly, observations were within 1\,day} multi-wavelength radio, mm, IR, optical and X-ray observations of \source\ as it softened through the HIMS and transitioned into the SIMS. To understand the full evolution of the jet and how it is connected to the accretion flow, we model the broad-band spectral evolution of this source. For the first time, we have been able to track the time-evolution of the jet spectral break during the rise of an outburst. Our results show a sudden decrease in \nub\ close to the HIMS to SIMS transition, indicating the onset of rapid compact jet quenching. We discuss the physical implications that arise from this rapid evolution.

\section{Observations}
\subsection{Radio observations}
\label{sec:radio}
\source{} was monitored in the radio band by the Australia Telescope Compact Array (ATCA) throughout its 2017/2018 outburst and re-brightenings (project codes C2601 and C3057, PI: Russell). See \citet{2019ApJ...883..198R} and \citet{2019ApJ...878L..28P} for the full details of the complete ATCA monitoring of the source during this outburst. In this paper, we use only the radio observations that had quasi-simultaneous sub-mm, IR, and optical observations to infer the evolution of the jet spectral break as the source evolved. Therefore, we discuss six ATCA observations taken between 2017 September 12 and 2017 September 21 (MJD~58008--58018), during the outburst rise. All observations were taken at central frequencies of either 5.5 and 9.0\,GHz, 17.0 and 19.0 GHz, or at all four frequencies, where each frequency pair (5.5/9 or 17/19 GHz) were recorded simultaneously. Errors on the absolute flux density scale include conservative systematic uncertainties of 4\% \citep[see, e.g.,][]{2010MNRAS.402.2403M,2016ApJ...821...61P}.

To explore rapid intra-observational source variability during the Sep 17 ATCA radio observation, we used \textsc{uvmultifit} \citep{2014A&A...563A.136M}. Fitting for a point source in the uv-plane, we explored flux density variations of the source down to 2-minute timescales, separating the 5.5 and 9\,GHz data into equally spaced 512-MHz sub-bands, and the 17 and 19\,GHz data into 1\,GHz sub-bands. For these data, uncertainties were estimated by treating the inner two 17/19\,GHz scans of the secondary calibrator as a target and comparing the flux densities of those `target' scans to the calibrator scans. For each of those two scans the flux density remained within 0.5\% of the expected value. Therefore, we apply a 0.5\% relative uncertainty to the intra-observational data on Sep 17. All radio data used in this work have been tabulated in the Appendix, Table~\ref{tab:data}. 

\subsection{Millimetre/Sub-millimetre observations}
\label{sec:alma}
The Atacama Large Millimetre/Sub-Millimetre Array (ALMA) observed \source\ (PI: Tetarenko, project code: 2016.1.00925.T) on 2017 Sep 11 (MJD 58007.9017--58007.9711) and Sep 21/22 (MJD 58017.9296--58018.0319 and MJD 58018.9272--58018.9641). Data were taken sequentially in Bands 3, 4, and 6, at central frequencies of 97.5, 145, and 236\,GHz, respectively. The ALMA correlator was set up to yield $4\times2$ GHz wide base-bands. During our observations, the 12m array was in its Cycle 4 C8 configuration, with between 39--45 antennas, spending $\sim$9.8/12.1/18.9\,min total on the target source in Bands 3, 4, and 6. The median precipitable water vapour (PWV) during the observations was 1.01\,mm on Sep 11 and 1.26\,mm on Sep 21/22. We reduced and imaged the data with the Common Astronomy Software Application package (\textsc{casa}, version 5.1.1; \citealt{2007ASPC..376..127M}), using standard procedures outlined in the \textsc{casa}Guides for ALMA data reduction\footnote{\url{https://casaguides.nrao.edu/index.php/ALMAguides}}. We used J1617--5848, J1427--4206, and J2056--4714 as bandpass \& flux calibrators, and J1631--5256 as a phase calibrator for all the observations. To image the continuum emission, we performed multi-frequency synthesis imaging on the data with the \texttt{tclean} task within \textsc{casa}, with natural weighting to maximize sensitivity. Multiple rounds of phase only self-calibration were implemented, down to solution intervals of 20\,seconds. We measured flux densities of the source by fitting a point source in the image plane (with the \texttt{imfit} task). 

 For the Sep 12 and Sep 21 ALMA data, the sub-mm emission was variable during the observation. This variability arises from short timescale brightening and fading of the flat-spectrum compact jet. To account for the full range of flux densities observed, we applied errors of $\pm$10, 15, and 20\,mJy to the 97.5, 145, and 236\,GHz, ALMA data on Sep 12, respectively. On Sep 21, we applied errors of 10 and 3\,mJy to the 97.5 and 145\,GHz data, respectively. These errors are larger than the expected 5\% uncertainty for ALMA bands $<$350\,GHz\footnote{\url{https://almascience.eso.org/documents-and-tools/latest/documents-and-tools/cycle8/alma-technical-handbook}}. All ALMA mm/sub-mm fluxes are tabulated in the Appendix, Table~\ref{tab:data}.

\subsection{Near-IR and optical monitoring}
\citet{2018ApJ...867..114B} presented and discussed dense optical, near-IR, and mid-IR monitoring of \source\ during the rise phase of its outburst. In this paper, we take optical and IR observations that were close in time (typically $\pm$0.5\,day) to our radio and sub-mm observations. All IR and optical data were de-reddened with an \nh\ of (3.84$\pm$0.03)$\times$10$^{22}$\,cm$^{-2}$ (see \citealt{2018ApJ...867..114B} for full details). Data used are provided in the Appendix, Table~\ref{tab:data}.

\begin{table*}
\caption{Best fitting parameters from the broad-band radio-to-X-ray modelling of \source\ (Figure~\ref{fig:jetbreak}). The Sep 12 -- 17 epochs were taken during the HIMS, while the source was in the SIMS on Sep 21 \citep{2018MNRAS.480.4443T}.  $\alpha_{\mathrm{thick}}$ is the spectral index of the optically-thick synchrotron emission, while $\alpha_{\mathrm{thin}}$ is the index of the optically-thin synchrotron emission. \nub\ is the frequency of the jet spectral break, $kT_{\mathrm{disk}}$ is the disk temperature, $\Gamma$ is the photon index of the high energy X-ray emission, $L_{\rm c}/L_{\rm d}$ is the fraction of the luminosity emitted in the corona to the total accretion flow luminosity. ${f_{\rm out}}$, the fraction of disk luminosity reprocessed in the outer disk, was tied across all epochs providing a best fit value of ($1.6^{+2.9}_{-0.8}$)$\times$10$^{-7}$. \nh, the line of sight equivalent hydrogen column density, was also tied across all epochs, giving a best fit value of $5.15^{+0.03}_{-0.02}\times10^{22}$ cm$^{-2}$. All quoted uncertainties are $1$-$\sigma$. The fits are shown in Figure~\ref{fig:jetbreak}. The best-fit statistic is $\chi^{2}/\rm{d.o.f.} = 2785.075/1490 = 1.87$.}
\footnotesize
\begin{tabular}{ccccccccc}
\hline
 Date &  $\alpha_{\mathrm{thick}}$ & $\nu_{\mathrm{break}}$ &  ${\alpha_{\mathrm{thin}}}^{\rm a}$  & \texttt{BPL} & $kT_{\mathrm{disk}}$ & $\Gamma$ &  $L_{\rm c}/L_{\rm d}$ & \texttt{diskir}  \\
 & & (Hz) & & norm.  & (keV) & &  & norm.  \\ 
 & & & & $10^{3}$ & & & & 10$^6$ \\ 

\hline

2017 Sep 12 & $0.17 \pm 0.02$ & $(8.6^{+2.6}_{-2.2}) \times 10^{12}$ & $-0.83 \pm 0.09$ & $4.29^{+1.29}_{-0.84}$ & $0.234 \pm 0.004$ & $1.74^{+0.01}_{-0.02}$ & $5.7^{+0.5}_{-0.4}$ &  $1.16^{+0.07}_{-0.06}$ \\ 

2017 Sep 14 & $0.09^{+0.02}_{-0.01}$ & $(1.6^{+0.5}_{-0.4}) \times 10^{13}$ & $-0.83 \pm 0.09$ & $1.31^{+0.40}_{-0.27}$ & $0.245^{+0.003}_{-0.002}$ & $1.93^{+0.01}_{-0.02}$ & $1.45^{+0.09}_{-0.07}$ &  $2.9 \pm 0.2$ \\

2017 Sep 15 & $0.17 \pm 0.02$ & $(7.2^{+1.9}_{-1.7}) \times 10^{12}$ & $-0.83 \pm 0.09$ & $4.66^{+1.72}_{-1.04}$ & $0.236 \pm 0.003$ & $1.96^{+0.01}_{-0.02}$ & $1.39^{+0.11}_{-0.09}$ &  $3.3 \pm 0.3$ \\

2017 Sep 16 & $-0.010 \pm 0.005$ & $(3.4 \pm 0.7) \times 10^{13}$ & $-0.83 \pm 0.09$ & $0.22 \pm 0.02$ & $0.244^{+0.003}_{-0.002}$ & $1.86 \pm 0.02$ & $2.7 \pm 0.2 $ &  $1.70^{+0.13}_{-0.12}$ \\

2017 Sep 17$^{\rm b}$ & $0.24^{+0.01}_{-0.02}$ & $(7.9^{+1.0}_{-1.4}) \times 10^9$  & $-0.26^{+0.06}_{-0.05}$ & $14.1^{+3.4}_{-3.3}$ & $0.241^{+0.003}_{-0.002}$ & $1.95 \pm 0.01$ & $1.75^{+0.10}_{-0.09}$ &  $4.0^{+0.4}_{-0.3}$ \\

2017 Sep 21$^{\rm c}$ & -- & $\leq 4.5 \times 10^9$ & $-0.52 \pm 0.02$ & $37.0^{+0.4}_{-0.3}$ & $0.09^{+0.03}_{-0.07}$ & $2.17 \pm 0.01$ & $9.3^{+0.7}_{-2.5}$ &  $55.0^{+1.5}_{-2.0}$ \\

\hline
\end{tabular}
\label{tab:pars}
\begin{flushleft} 
\vspace{-2mm}
$^{a}$ Parameter tied across the first 4 epochs due to lack of IR data on Sep 12 and Sep 14. \\
$^{b}$ For this epoch, $\alpha_{\rm thick}$, \nub, $\alpha_{\rm thin}$ were determined from our detailed radio analysis (see Section~\ref{sec:variability}), although we note that leaving these parameters free in our broadband modelling produced similar results. \\
$^{c}$ For this final multiwavelength epoch, the radio and sub-mm emission originated from optically-thin synchrotron emission from the transient jet \citep{2019ApJ...883..198R}. 
\end{flushleft}
\end{table*}

\subsection{X-ray observations}
\label{sec:x-ray}
The Neil Gehrels \swift{} Observatory X-ray telescope (XRT) monitored \source\ every $\sim$1--2\,days during the rise of the 2017 outburst. For full details of the \swift-XRT monitoring, see \citet{2018MNRAS.480.4443T}. In this work, we used only the observations closest in time to our ATCA monitoring ($\pm$0.5\,day). \swift-XRT data were prepared, extracted and analyzed with the standard tools in \textsc{heasoft} (version 6.25). We ran the \texttt{xrtpipeline} to apply the newest calibration. The target source was extracted with \texttt{XSELECT} (version 2.4) with a box along the WT readout strip with a length of 35 (82.506$^{\prime\prime}$). The inner 20 pixels (47.146$^{\prime\prime}$) were excluded in order to eliminate pile-up. The background was extracted with an annulus region centered on the source, with radii of 46 pixels (108.436$^{\prime\prime}$) and 100 pixels ($235.731^{\prime\prime}$).

We note that there appeared to be some discrepancy between spectral results from different X-ray telescopes, in particular an offset in the slope of the X-ray photon index and normalisation between \swift-XRT, \textit{NuSTAR} (Nuclear Spectroscopic Telescope Array) and \textit{XMM-Newton} X-ray telescopes (for further discussions see also, e.g., \citealt{2014MNRAS.437..316K,2016ApJ...824...37L,2016MNRAS.461.1967I,2017MNRAS.466.2910S,2017MNRAS.466L..98V}). However, the evolution trend of each parameter was the same regardless of the X-ray telescope. Therefore, as we focus on relative changes to the accretion flow properties, our key findings remain unchanged across all X-ray telescopes. Due to its high observing cadence and more complete monitoring, in this work we used X-ray observations taken with \swift-XRT.

\section{Broad-band spectral modelling}

\begin{figure*}
\centering
\includegraphics[width=0.95\textwidth]{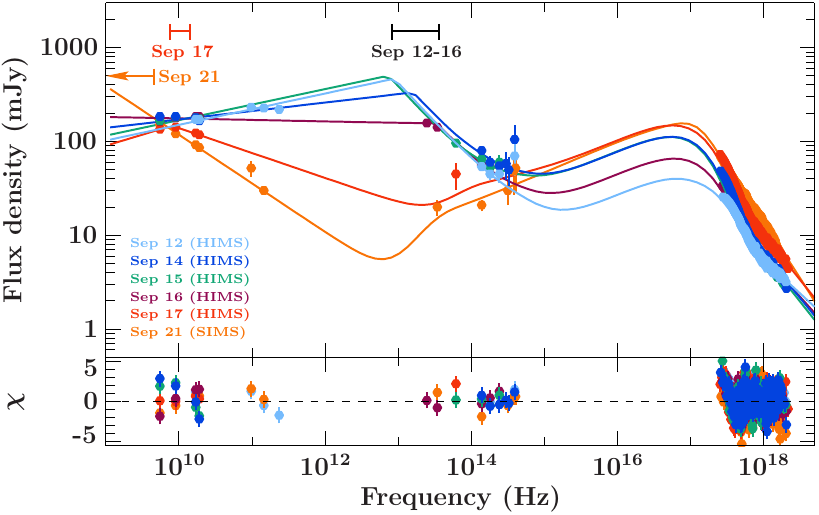}
\caption{Broad-band multi-wavelength modelling of \source\ during the rise phase of its 2017 outburst (as the source transitioned from the HIMS to the SIMS; states identified in the legend). Solid lines represent the broad-band models for each epoch where different colours depict different observational epochs. The horizontal bars at the top of the figure show the full ranges of \nub\ for the given dates. Fit residuals are shown in the lower panel. Parameters are provided in Table~\ref{tab:pars}. This plot highlights the rapid change to the jet around Sep 17, after it had remained relatively steady for the previous $\sim$5\,days. On Sep 17, however, \nub\ rapidly decreased by $\sim$3 orders of magnitude in frequency, residing within the radio band. For plotting purposes only, the plotted X-ray data have been de-absorbed.} 
\label{fig:jetbreak}
\end{figure*}

For the six epochs with quasi-simultaneous, multi-wavelength coverage, the broad-band spectral energy distribution (SED) was modelled with the Interactive Spectral Interpretation System (\textsc{isis}, version 1.6.2-35; \citealt{2000ASPC..216..591H}). We fit the radio to IR data with a broken power-law (representing the optically thick and thin synchrotron emission from the compact jet); we include an artificial exponential cutoff so as not to assume that the optically-thin synchrotron emission from the jet contributes to the X-ray emission. The X-ray data were modelled with an absorbed irradiated inner and outer disk \citep{diskir}. The full multi-wavelength model is \texttt{tbnew*(highecutoff*bknpower+diskir)}.

The data were fit with the \textsc{isis} implementation of a Markov chain Monte Carlo algorithm (MCMC), based on the python package by \cite{MCMC}. We use 20 walkers per free parameter and evolve the chain for 10000 loops, taking the first 4000 to be the ``burn-in'' period to ensure convergence of the algorithm. We found that this was roughly the time required for the chain to converge, which we define as the point past which the posterior distribution of the parameters stops evolving, and the acceptance rate of the chain stabilizes. We define the best fitting values as the median of the one-dimensional posterior distribution, and the $1$-$\sigma$ uncertainties as the intervals of the posterior distribution in which 68\% of the walkers are found. Fits are shown in Figure~\ref{fig:jetbreak}, with best-fit parameters given in Table~\ref{tab:pars}.

The broad-band data from each epoch were fit independently, with the exception of the optically-thin jet spectral index, $\alpha_{\rm thin}$, which was tied across the epochs from September 12 to 16, and the fraction of disk luminosity reprocessed in the outer disk, $f_{\rm out}$, which was tied across all epochs. Leaving it free did not improve the quality of the fit, nor change the best-fitting parameter values significantly. We chose to tie these parameters in order to reduce the model's inherent degeneracy, as both of these parameters set the ratio of emission from the jet or accretion flow to the optical/IR emission on Sep12 -- 16. \nh\ was free, but tied across all epochs providing a result similar to previously determined values from more comprehensive X-ray studies of this source \citep[e.g.,][]{2018MNRAS.480.4443T}. The temperature of the corona was fixed at $T_{\rm e}=100\,\rm{keV}$. The remaining parameters of \texttt{diskir} were frozen to $\rm{R_{\rm irr}}=1.2$, $\rm{f_{\rm in}}=0.1$, $\log(R_{\rm out})=4.5$\footnote{${\rm R_{irr}}$ is the radius of the illuminated disk in terms of the inner disk radius. ${\rm f_{in}}$ is the fraction of the Compton luminosity thermalised in the inner disk, and $\log(R_{\rm out})$ is the log of the outer disk radius in terms of the inner disk radius. The set values are typical values for these parameters.}. In principle, the Compton component from \texttt{diskir} can be made up of thermal electrons, non-thermal electrons, or synchrotron self-Compton (due to the presence of a magnetic field). In this work, we do not attempt to distinguish or identify the contribution from different Compton components. Additionally, due to the high line of sight absorption, the companion was not detected. Therefore, the IR and optical excess, which exceeded the jet contribution, is due to irradiation of the inner and outer accretion disk.

\begin{figure}
\centering
\includegraphics[width=1.0\columnwidth]{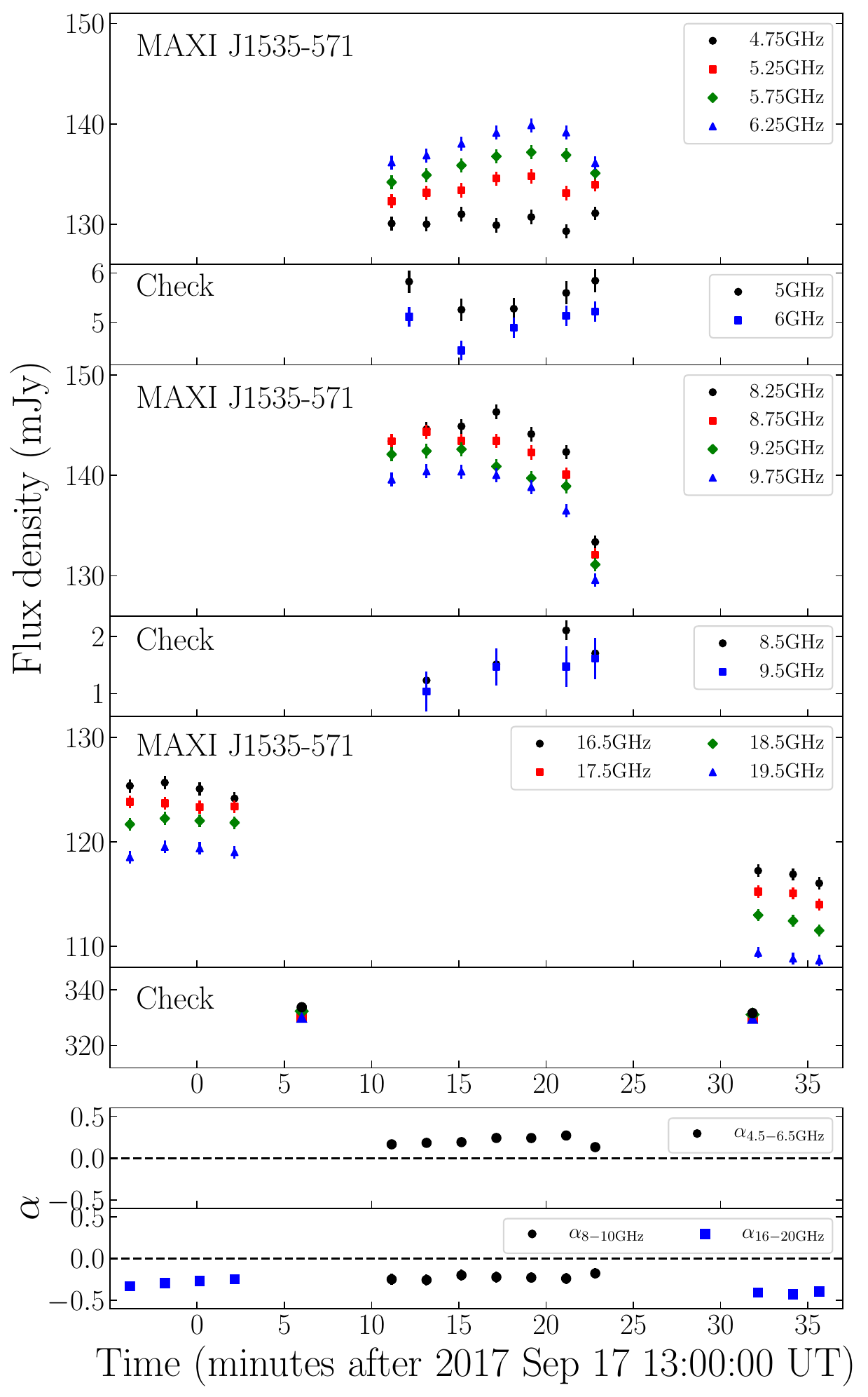}
\caption{Intra-observation variability of \source\ on 2017 Sep 17, where each frequency band was separated into finer frequency sub-bands. To show that the observed changes (both in time and frequency) were intrinsic to the source and not instrumental, we show the emission from a comparison/check source. (\textit{Top panel:}) shows the target flux densities recorded in the 5.5\,GHz band, which was separated into 500\,MHz sub-bands. (\textit{Second panel:}) provides the 5.5\,GHz-band flux densities of a nearby field source, which was broken into 1\,GHz sub-bands (due to the lower source brightness than the target). (\textit{Third} and \textit{Fourth panels:}) same as the first two panels but for the 9\,GHz observations, respectively. (\textit{Fifth panel:}) presents the 17 and 19\,GHz radio observing bands separated into 1\,GHz sub-bands. (\textit{Sixth panel:}) provides the flux densities of two scans of the phase calibrator when treated as the target, where the symbols and colours are the same as the fifth panel. (\textit{Seventh} and \textit{Eighth panel:}) show the spectral index for the 4.5--6.5\,GHz and $>$8\,GHz radio data, respectively. The intra-band spectral and time variability implies that there was a spectral break within the radio band during this observation, and that break was decreasing in frequency over time}.
\label{fig:inradio}
\end{figure}

\section{Results}
\subsection{Source evolution}
During the first epoch of multi-wavelength monitoring (on 2017 Sep 12), \source\ was in the HIMS with a relatively hard X-ray spectrum ($\Gamma = 1.74^{+0.01}_{-0.02}$) and a flat to slightly inverted radio spectrum (Table~\ref{tab:pars}). The radio, mm, and IR broad-band data indicated that during this epoch, the optically-thick synchrotron emission extended beyond the mm band, such that \nub$\sim 10^{13}$\,Hz.

Our next four multi-wavelength epochs were taken on consecutive days, running from 2017 Sep 14 to 2017 Sep 17. For all of these, the source remained within the HIMS. While the X-ray spectrum had softened from the first multi-wavelength epoch, during these four consecutive observations the X-ray photon index remained relatively steady ($\Gamma \approx 1.85 - 1.95$), and, aside from the first epoch, the disk temperature was also stable (where $kT_{\mathrm{disk}} \approx 0.24$\,keV). Between Sep 14 and Sep 16 the compact jet appeared steady, with $\alpha_{\mathrm{thick}}$ remaining flat to slightly inverted and \nub\ around $\sim$10$^{13}$\,Hz. However, for the Sep 17 epoch, we observed a sudden and dramatic change in the jet emission. Our observations showed \nub\ had decreased to $(7.9^{+1.0}_{-1.4})\times 10^{9}$\,Hz, lying within the radio band with evidence that it was decreasing throughout the radio observation (see Section~\ref{sec:variability}).

For our final multi-wavelength epoch on 2017 Sep 21, \source\ had transitioned into the SIMS \citep{2018MNRAS.480.4443T}. The X-ray spectrum had softened ($\Gamma = 2.17\pm0.01$). During this epoch, we detect only steep spectrum radio emission originating from a discrete ejection associated with the transient jet (see \citealt{2019ApJ...883..198R} for further details).

\subsection{\nub\ within the radio band on 2017 Sep 17}
\label{sec:variability}

The radio spectrum on 2017 Sep 17 was not well represented by a single power law. To investigate this further we performed a more detailed timing and frequency analysis of these radio data, where we determined the flux densities of the target on 2-minute time-intervals, and on finer frequency scales (512\,MHz sub-bands for the 5.5 and 9\,GHz data, and 1\,GHz sub-bands for the 17 and 19\,GHz radio data; see Appendix~\ref{App:Sep17} for the full table of results from the short-time interval study). This analysis shows that during our Sep 17 radio observation the 4.5--6.5\,GHz radio emission was inverted, such that $\alpha_{\mathrm{4.5\,GHz}}^{\mathrm{6.5\,GHz}} \approx 0.2$ (Figure~\ref{fig:inradio}). While the 4.5--6.5\,GHz emission was time-variable, the variability did not show any clear trends in time or across different frequency bands. At the same time, the 8--10\,GHz emission exhibited a steep radio spectrum ($\alpha \approx -0.27$) and the emission appeared to fade towards the end of the 15\,min observation (from $\approx$143\,mJy to $\approx$130\,mJy). The steep spectrum 16--20\,GHz emission appeared steady for the first part of the observation, but was observed to be fading steadily during the second part. To check that the variability and observed radio spectrum were intrinsic to the source, for the 4.5--6.5 and 8--10\,GHz observations we compared our results to a comparison source that was within the field\footnote{The nearby field source is located at Right Ascension: 15$^{\rm h}$35$^{\rm m}$11.462$^{\rm s}$ and Declination: -57$^{\rm d}$10$^{\rm m}$46.37$^{\rm s}$ ($\approx$160\arcsecond\ to the North North West of \source).} and did not detect similar behaviour. For the 16--20\,GHz observations, we did not detect this check source, or any others in the field. Therefore, to test if the observed variability was intrinsic to \source\, we re-calibrated the data treating the inner two scans of the phase calibrator as target scans. As shown in Figure~\ref{fig:inradio} these two `target' scans remained steady, indicating that the variability was related to \source (see Section~\ref{sec:radio} for further details). The shape of the radio spectrum and behaviour of the radio emission during our Sep 17 ATCA observation implied that the jet spectral break was within or between the 4.5-6.5 and 8--10\,GHz observing bands and decreasing in frequency during the Sep 17 radio observation (Figure~\ref{fig:bknpower} and Table~\ref{tab:inradio}).

There was a mid-IR VISIR (J8.9-band, 3.44$\times$10$^{13}$\,Hz) observation taken $\approx$12\,hours before our radio observation on 2017 Sep 17 (observed between 00:52 and 01:27 UT; see \citealt{2018ApJ...867..114B}). This bright mid-IR detection had a de-reddened integrated flux density of 141$\pm$12\,mJy, which indicated that the jet was still bright at mid-IR frequencies ($\sim 3 \times 10^{13}$\,Hz) at this time. However, this mid-IR emission faded from $\sim$180\,mJy to $\sim$110\,mJy over the 35\,minute observation (see figure 3 in \citealt{2018ApJ...867..114B}), implying that the IR and optical jet emission was rapidly fading. As a check, assuming that this mid-IR emission was decaying exponentially, fitting and extrapolating the VISIR flux densities to the time of our radio observation suggests that we would expect no jet contribution at mid-IR frequencies by the time of our radio observation. Therefore, we do not include this mid-IR data point in our multiwavelength fit for the Sep 17 data, instead we chose to include it in our Sep 16 data, when the compact jet was still on. \citet{2018ApJ...867..114B} reported a 45$\pm$5\,mJy de-reddened M-band (6.19$\times$10$^{13}$\,Hz) detection of the source $\sim$11\,hours after our Sep 17 radio epoch. While the detected IR emission at this epoch had faded considerably, there may be some contribution from a jet. However, the authors suggest that the observed IR variations detected appear to be associated with an intermittent jet or flaring from the jet base, and not the steady compact jet detected in the days before (see \citealt{2018ApJ...867..114B} for further discussion). These two mid-IR observations taken $\sim$11--12\,hours before and after our radio observation show the jet had faded at IR frequencies over this time, in agreement with our radio results, although the jet base may have been flaring. 

We note that the rapid radio variability during our 2017 Sep 17 epoch will have had some effect on our SED modelling results. However, the key finding of the break residing in the radio band remains.

\section{Discussion}

With our multi-wavelength observations of \source\ we detected the rapid evolution of the compact jet during the HIMS as the source transitioned to the SIMS. In particular, our observations show \nub\ decreased suddenly as the source transitioned from the hard to soft states. In this section, we discuss the implications of this finding for how the compact jet emission is believed to quench.

\subsection{Time-evolution of \nub}

Previously, the time-evolution of \nub\ during the rise-phase of an outburst has only been implied by the change from an optically-thick to optically-thin radio spectrum, or the rate at which the IR emission fades as the source transits from the hard state to the soft state (through the intermediate states). IR rates of decay have suggested an evolution of $\sim$1--2\,weeks (see \citealt{2001ApJ...554L.181J,2019ApJ...887...21S}). With our radio-to-IR coverage, we have been able to accurately track the evolution of \nub\ as the source moved from the HIMS to the SIMS. Our observations show that \nub\ decreased in frequency by $\sim$3 orders of magnitude (evolving from $\sim$10$^{13}$\,Hz to $\sim$10$^{10}$\,Hz) in $<$24\,hours (Table~\ref{tab:pars} and Figure~\ref{fig:jetbreak}). In fact, the bright, jet dominated mid-IR detection $\sim$12\,hours before our Sep 17 radio observation suggest that the jet break evolved from the IR to radio band in $\sim$12\,hours or less, implying a minimum rate of $\sim$1-order of magnitude in frequency in $<$4 hours. However, detailed time and frequency analysis of our Sep 17 radio observation shows \nub\ within the radio band, where we were able to place more stringent constraints, estimating \nub\ to be decreasing at a rate of $\sim$1.8\,GHz over the $\sim$15\,min observation (which approximates to one order of magnitude in 1.4\,hours assuming it was decreasing linearly with time).

\begin{figure}
\centering
\includegraphics[width=0.94\columnwidth]{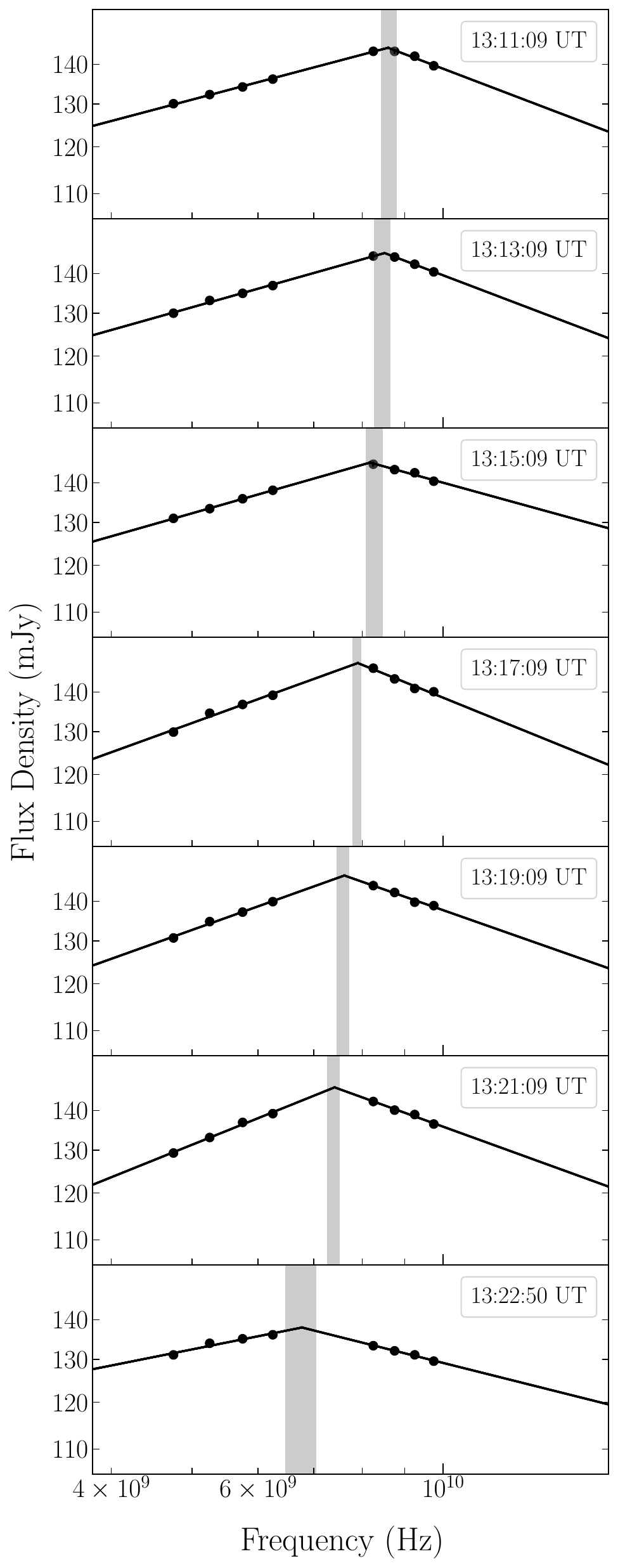}
\caption{Broken power-law fits of the ATCA radio emission from \source\ on 2017 Sep 17 (errors on each data point are shown, but in most cases are smaller than the marker size). Here we show a time-series of each successive 2-minute radio snapshot, where the top panel indicates the earliest epoch and the lower panels show its progressive evolution. The mid-point time (in UT) of each snapshot is provided in each panel. The grey shaded region on each panel represents the best fit \nub\, and its uncertainties. These results imply that the jet spectral break was within the radio band during our Sep 17 ATCA observation, and they suggest that the frequency of the jet break was decreasing during the $\sim$15\,minute radio observation.}
\label{fig:bknpower}
\end{figure}

To estimate the rate at which \nub\ was evolving within the radio, within \textsc{ISIS}, we fit a broken power-law to each 2-min time segment of the strictly simultaneous 4.5--10\,GHz data (Figure~\ref{fig:bknpower} and Table~\ref{tab:inradio}). The changes in the radio data imply that \nub\ decreased by 1.83$^{+0.50}_{-0.45}$\,GHz over the 15\,min observation (Figure~\ref{fig:bknpower} and Table~\ref{tab:inradio}).

\begin{table}
\centering
\caption{Best fitting parameters from the broken power-law fits of the 2-minute interval data from the 2017 Sep 17 ATCA radio data (shown in Figure~\ref{fig:bknpower}). Here we report the spectral indices and \nub\ for each time segment. }
\footnotesize
\begin{tabular}{ccccc}
\hline
\multicolumn{2}{c}{Epoch} &  $\alpha_{\mathrm{thick}}$ & $\nu_{\mathrm{break}}$ &  ${\alpha_{\mathrm{thin}}}$  \\ 
MJD & Time & &  & \\
($\pm 0.0007$) & (UT) & & (GHz) & \\

\hline

58013.5494 & 13:11:09 & 0.18$\pm$0.01 & 8.6$\pm$0.2 & -0.25$\pm$0.05 \\
58013.5508 & 13:13:09 & 0.19$\pm$0.01 & 8.52$\pm$0.15 & -0.26$\pm$0.05 \\
58013.5522 & 13:15:09 & 0.19$\pm$0.01 & 8.22$^{+0.25}_{-0.15}$ & -0.19$\pm$0.04 \\
58013.5536 & 13:17:09 & 0.24$\pm$0.02 & 7.90$\pm$0.1 & -0.27$\pm$0.03 \\
58013.5550 & 13:19:09 & 0.24$\pm$0.02 & 7.60$^{+0.10}_{-0.15}$ & -0.23$\pm$0.02 \\
58013.5564 & 13:21:09 & 0.27$\pm$0.02 & 7.40$\pm$0.15 & -0.24$\pm$0.02 \\
58013.5575 & 13:22:50 & 0.13$\pm$0.02 & 6.77$^{+0.25}_{-0.30}$ & -0.17$\pm$0.02 \\

\hline
\end{tabular}
\label{tab:inradio}
\end{table}

Recent studies have inferred rapid changes of \nub\ in two other BH candidates, MAXI~J1659$-$152 \citep{2013MNRAS.436.2625V} and GX~339$-$4 \citep{2011ApJ...740L..13G}. In the 2010 outburst of MAXI~J1659$-$152, the jet spectral break was found to be intermittently in the observed radio bands and at (unobserved) higher frequencies, as the source switched back and forth between the SIMS and HIMS, before settling into the soft state. In one such oscillation, it was possible to infer that \nub\ decreased by at least one order of magnitude in less than 3\,days \citep{2013MNRAS.436.2625V}. However, \nub\ could not be more precisely constrained at all the epochs in which it fell above the radio bands, because of the lack of mm and IR observations. For GX~339$-$4 the IR spectrum in the 10$^{13}$--10$^{14}$\,GHz band changed from optically thick to optically thin between two Wide-Field Infrared Survey Explorer (WISE) observations taken 1.58\,hours apart, in 2010 \citep{2011ApJ...740L..13G}. This spectral change implies that \nub\ decreased by more than one order of magnitude in $<$1.58\,hours, but in that case, too, it was not possible to constrain the rate more precisely because there were no observations above and below the WISE band.

With our multi-band study of \source, we have identified an extremely rapid rate of change of \nub, equivalent to one order of magnitude in $\sim$1.4\,hours, and a total of three orders of magnitude in $<$1 day. Most importantly, we have been able to, for the first time, pinpoint its location before, during, and after such a rapid transition.

\subsection{Radius and magnetic field of the particle acceleration zone}

\begin{figure}
\centering
\includegraphics[width=0.99\columnwidth]{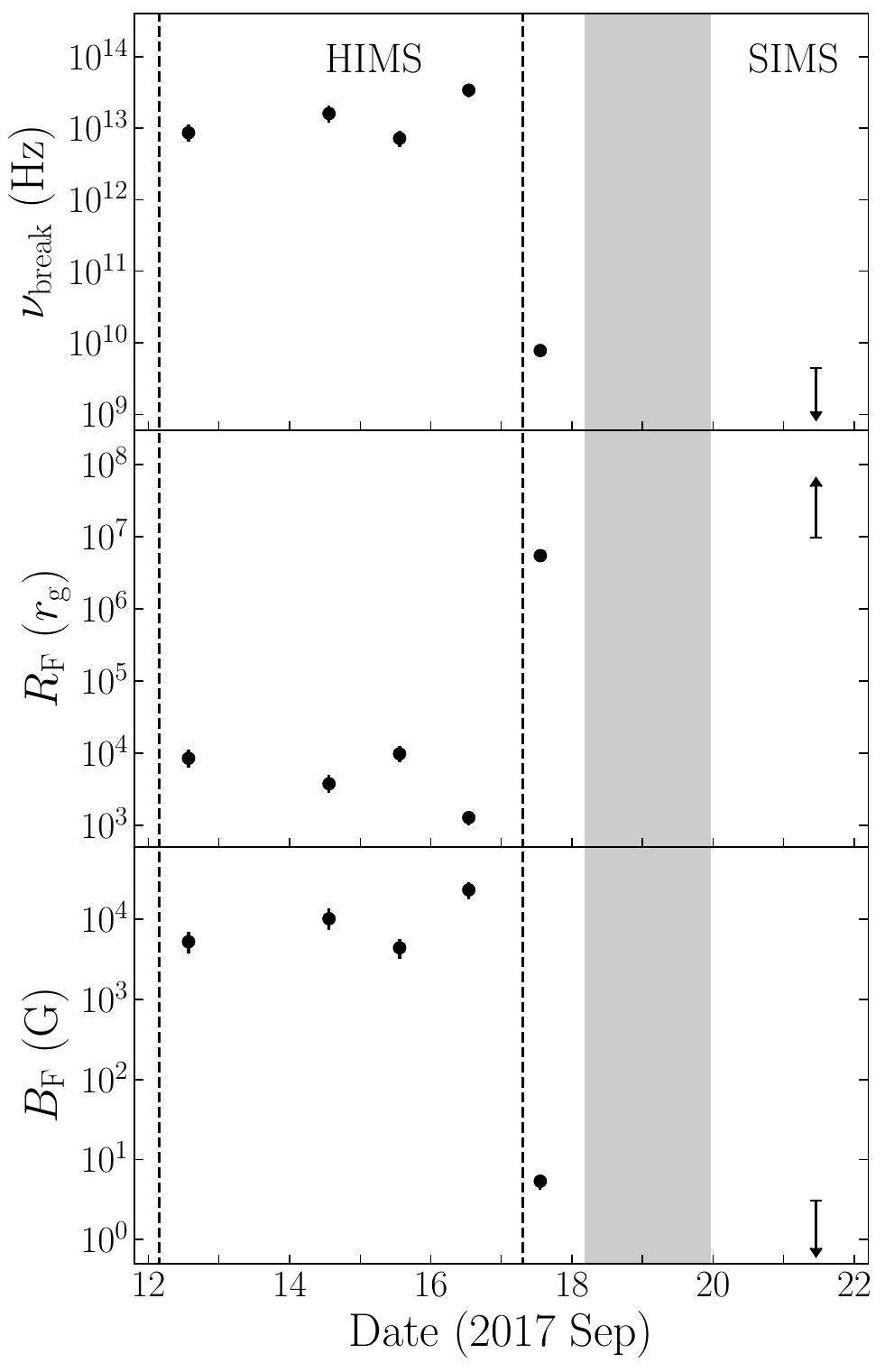}
\caption{Evolution of \nub\ and the subsequent changes to \rz\ and \bz\ during the rise-phase of \source's 2017 outburst. \textit{Top panel:} The frequency of the jet spectral break. \textit{Middle panel:} Radius of the first acceleration zone, \rz, in gravitational radii (assuming a 10\,M$_\odot$ black hole). \textit{Lower panel:} The magnetic field at the jet base, \bz. The grey region marks the HIMS$\rightarrow$SIMS transition, where the expanse is the uncertainty in the time that it occurred \citep{2018MNRAS.480.4443T}. The vertical dashed lines show the full range estimated for the ejection of the transient jet knot detected from the system \citep[see][]{2019ApJ...883..198R}. \nub\ shifted by $\sim$3 orders of magnitude in $\leq$1\,day, resulting in rapid changes to the radius and magnetic field of the acceleration region.}
\label{fig:rfaz}
\end{figure}

As outlined by \citet{2011A&A...529A...3C}, following \citet{1979rpa..book.....R} and \citet{2011hea..book.....L}, with the frequency and flux density of the jet spectral break we can estimate the radius (\rz) and magnetic field strength (\bz) of the first acceleration zone, which is where particle acceleration begins (Figure~\ref{fig:rfaz}). Assuming equipartition between the particle energy and magnetic field energy density, \bz\ and \rz\ can be approximated using the frequency and flux density of the spectral break, such that: 

\begin{equation}
B_{\rm F} \propto S_{\nu,{\rm b}}^{-2/(2p+13)}  \nu_{\rm b}
\label{eq:B_F_prop}
\end{equation}
and
\begin{equation}
R_{\rm F} \propto  S_{\nu,{\rm b}}^{(p+6)/(2p+13)} \nu_{\rm b}^{-1}, 
\label{eq:RF_full}
\end{equation}
where $\nu_{\rm b}$ is the break frequency, $S_{\nu,{\rm b}}$ is the flux density at the break frequency, and $p$ is the slope of the electron energy spectrum, such that $p=1-2\alpha_{\rm thin}$. See Appendix~\ref{App:chaty} for the full equations.

From Equations~\ref{eq:B_F_prop} and \ref{eq:RF_full}, our broad-band modelling implies that during the HIMS when the compact jet appeared to be relatively steady, \rz\ $\sim$10$^3$--10$^4$\,\rg\ and \bz\ $\sim$10$^4$\,G (Figure~\ref{fig:rfaz}). Then, as the compact jet began to rapidly fade during the early stages of the jet being quenched, \rz\ and \bz\ changed significantly in the space of $\approx$1\,day. Between Sep 16 and 17, we estimate that \rz\ increased from $\sim10^{4}\,r_{\mathrm{g}}$ to $\sim10^{7}\,r_{\mathrm{g}}$ (assuming a 10\,M$_\odot$ BH). Over the same time, \bz\ decreased from $\sim10^{4}$\,G to $\sim10^{1}$\,G. 

MAXI~J1836$-$194 is the only other case where the time evolution of \rz\ and \bz\ has been determined \citep{2014MNRAS.439.1390R}, in that case during the outburst decay. In that system, the compact jet was observed to re-establish after the peak of the outburst; hence, while the values of \rz\ and \bz\ are comparable, their evolution was reversed. However, in MAXI~J1836$-$194 that evolution occurred over a period of $\sim$6\,weeks, as opposed to the $\lesssim$1\,day we determine for \source.

\subsubsection{Adiabatic and radiative cooling timescales}
\label{sec:timescales}

\begin{figure}
\centering
\includegraphics[width=0.99\columnwidth]{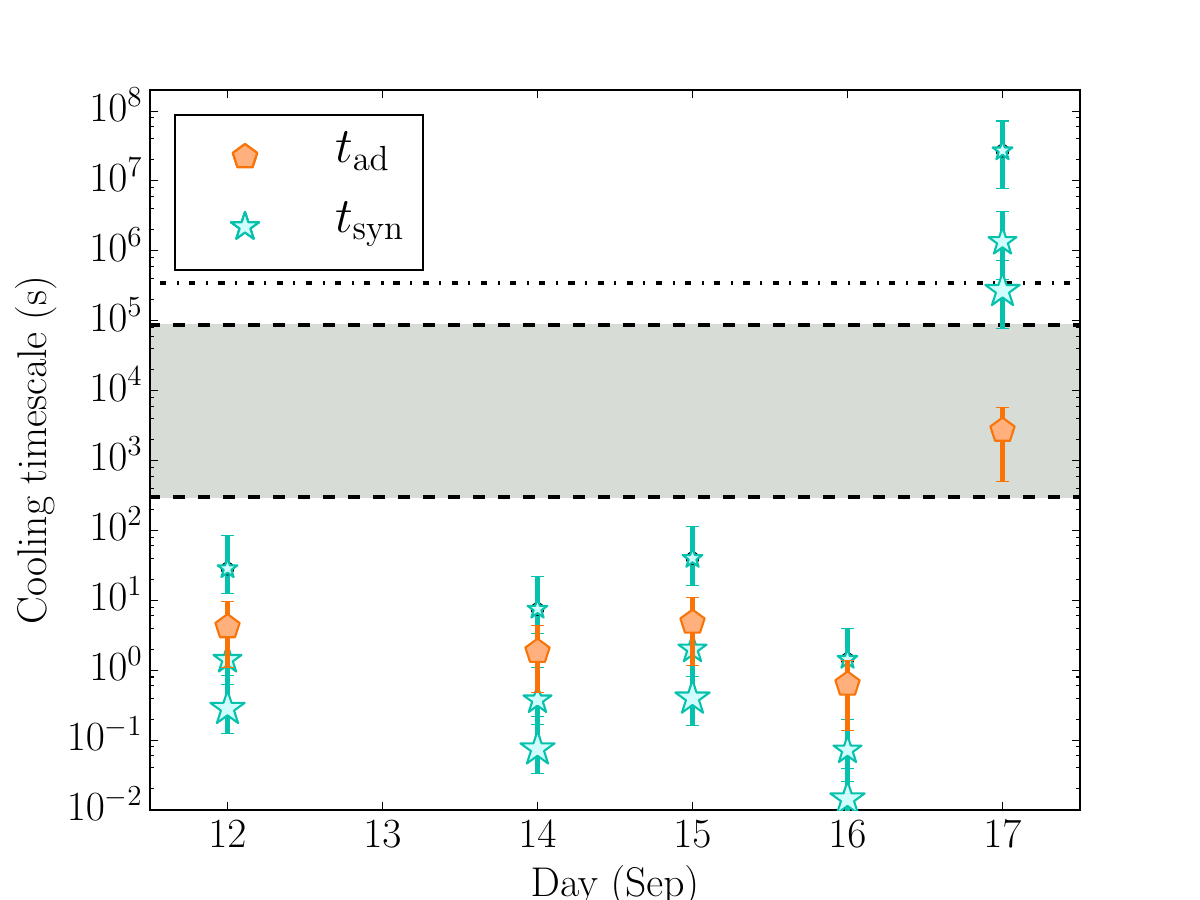}
\caption{Adiabatic and radiative cooling timescales of the compact jet emission from our Sep 12 -- 17 epochs. Orange pentagons indicate the adiabatic timescale during each epoch, cyan stars the radiative timescale for electrons with Lorentz factors 100, 20, and 1. Larger stars indicate larger electron Lorentz factors. The grey band represents the compact jet \nub\ variability timescales; the bottom dashed line corresponds the inter-observation variability timescale of a few minutes, the top dashed line corresponds to the time between the Sep 16 and 17 observations. The dot-dashed line corresponds to the time between the Sep 17 and 21 observations, which is the most conservative estimate for the rate at which the compact jet was quenched. The variation of \nub\ inferred from our observations implies that the acceleration region of the jet is moving away from the BH system on adiabatic/dynamical timescales.}
\label{fig:timescales}
\end{figure}

From the magnetic field and radius estimates (Figure~\ref{fig:rfaz}), we can estimate the cooling (adiabatic and radiative) timescales of the particles in the jet. Assuming a conical jet with a constant jet opening angle and expansion speed (of the ejected material) $\beta_{\mathrm{exp}}$, we define the adiabatic timescale as:
\begin{equation}
t_{\mathrm{ad}} =  \frac{R_{\rm F}}{\beta_{\mathrm{exp}} {\mathrm{c}}} = 0.48 \left(\frac{0.1}{\beta_{\mathrm{exp}}} \right) \left(\frac{R_{\rm F}}{10^{4}\,\rm{R_g}} \right) \left(\frac{M_{\mathrm{BH}}}{10\,M_{\odot}} \right) \,\rm{s}, 
\end{equation}
where $R_{\rm F}$ is the radius of the emitting region, $c$ is the speed of light, and $M_{\mathrm{BH}}/10M_\odot$ is the mass of the black hole in units of 10\,${\rm M_{\odot}}$. For a conical jet with a constant opening angle, the emission height above the BH $z_{\rm em} = R_{\rm F}/\beta_{\rm exp}$ (this can also be thought of as the light-travel time from the black hole to the emitting region including a beaming factor). For a given magnetic field $B_{\rm F}$ the synchrotron radiative timescale is:
\begin{equation}
t_{\mathrm{syn}} = \frac{3m_{\mathrm{e}}{\mathrm{c}}^{2}}{4\sigma_{\mathrm{t}}{\mathrm{c}}U_{\mathrm{b}}\gamma} = \frac{6\pi m_{\mathrm{e}} {\mathrm{c}}}{\sigma_{\mathrm{t}}B_{\rm F}^{2}\gamma} = 7.8 \left(\frac{10^4\,\rm{G}}{B_{\rm F}}\right)^{2}\left(\frac{1}{\gamma}\right) \,\rm{s},
\end{equation}
where $m_{\mathrm{e}}$ is the mass of the electron, $\sigma_{\mathrm{t}}$ is the Thomson cross section, $U_{\mathrm{b}} = B_{\rm F}^{2}/8\pi$ is the magnetic energy density, and $\gamma$ the electron Lorentz factor. We neglect inverse Compton losses for simplicity and due to lack of observational constraints, although we expect them to be small \citep{2004ApJ...600..368M}. Figure~\ref{fig:timescales} shows the evolution over all multi-wavelength epochs. On September 12 to 16, both adiabatic and radiative timescales are very short, as the optically-thin emitting region is very close to the black hole. However, this changes dramatically on Sep~17 as the break frequency shifts into the radio band. On this epoch, the time over which the compact jet was fading at higher frequencies was consistent with our estimates for the dynamical/adiabatic timescales, while the radiative timescales are longer by 2$-$3 orders of magnitude (even for modest electron Lorentz factors, $\gamma = 100$, where the Lorentz factors suggested by our calculated values of $B_{\rm F}$ are $\sim$0.56--56\footnote{Emitting at 9\,GHz.}). This result implies that the evolution of \nub, and hence, the high-energy quenching of the compact jet was not driven by some local phenomena, like the details of particle acceleration. Instead, a more likely explanation is some major change in the internal properties of the jet, such that the particle acceleration region suddenly shifted away with the jet flow, possibly disconnected from the accretion flow. 

On two separate occasions radio observations of the BH XRB GRS 1915$+$105 have shown a similarly rapid shutting off of the compact jet emission. \citet{2016ApJ...823...54P} analysed radio observations of the two events, taken $\sim$1-year apart with the Ryle Telescope. During both of these 15\,GHz radio observations the source was in its high plateau state, which is associated with the steady, compact jet. Separating the 4-hour long radio observations into 32\,s time intervals showed the radio emission fading rapidly from $\sim$100\,mJy and $\sim$70\,mJy down to a few mJy in 2003 April and 2004 April, respectively. The decay of the radio emission was attributed to the turning off, or rapid reduction in power, of the jet emission, creating a discontinuity propagating outwards along the jet \citep{2016ApJ...823...54P}. Such a scenario could result from the particle acceleration region rapidly shifting away from the central object, effectively switching off the compact jet emission in the same way our results suggest for \source.

\subsubsection{The optically-thin synchrotron spectrum}

The spectral indices we measured for the optically-thin synchrotron emission (Table~\ref{tab:pars}) are also suggestive of shorter synchrotron cooling timescales during the first four epochs (Sep 12 -- Sep 16), and longer for the Sep 17 epoch\footnote{Radio observations on Sep 21 show emission from the transient jet and not the compact jet.}. For the first four epochs, $\alpha_{\rm thin} \approx -0.83$, while on Sep 17 $\alpha_{\rm thin} \approx -0.3$. The steeper indices on Sep 12 -- Sep 16 may indicate the presence of a second spectral break arising from the rapid radiative cooling of the synchrotron emitting electrons in the jet spectrum at $\sim$IR frequencies (close to - or even below - the jet spectral break). This would result in the optically-thin spectrum appearing steeper by $\alpha \sim 1/2$. Then, as the jet emission began to switch off (on Sep 17) and the cooling timescales became much longer the frequency of the synchrotron cooling break, $\nu_{\rm cool}$), would be above the observed radio and sub-mm bands, such that $\alpha_{\rm thin}$ would appear shallower (consistent with un-cooled optically-thin synchrotron emission). Our understanding of the location and evolution of $\nu_{\rm cool}$ is not well constrained. Studies have suggested that it may lie in the UV-band during quiescence \citep[e.g.,][]{2013ApJ...773...59P}, at optical/IR frequencies in the hard state \citep{2014MNRAS.439.1390R}, before evolving into the X-ray band as the source transitions from the hard state to the soft state \citep[e.g.,][]{2012ApJ...753..177P,2013MNRAS.429..815R,2013MNRAS.434.2696S}. Therefore, with such poor constraints and suggested variability, it is plausible for the synchrotron cooling break to lie at IR frequencies during our Sep 12 -- Sep 16 observations, evolving to higher energies as the source transitioned towards the HIMS (on Sep 17 and beyond). Although we are not able to place any constraints on its specific frequency during the observations presented here.

 \subsection{Connection to the X-ray emission}

Interestingly, at the time that the jet dissipation region moved away, we did not detect any sudden and remarkable changes to the soft X-ray count rate, disk temperature, and X-ray photon index (Table~\ref{tab:pars}), although the radio event was rapid so we may have missed any clear but brief changes in the X-ray emission that occurred outside of the daily \swift\ X-ray observations. The closest-in-time \swift-XRT observation (which was used for our multi-wavalength SED fitting) was taken $\sim$10\,hours before we detected the jet spectral break within the radio band. If directly connected, we may expect that those X-ray observations may already have shown a change as the IR observations around the same time as this X-ray observation indicate that the high energy jet emission was already fading (see IR discussion in Section~\ref{sec:variability}, and discussions in \citealt{2018ApJ...867..114B}). We also compared our result to the \swift-XRT observation taken on Sep 18 \citep{2018MNRAS.480.4443T}, finding that while the disk temperature remained similar, the X-ray photon index did soften from $\Gamma$=1.87$^{+0.02}_{-0.01}$ on Sep 17 to $\Gamma$=2.17$^{+0.06}_{-0.07}$ on Sep 18 (results from \citealt{2018MNRAS.480.4443T}, to remain consistent) and the soft X-ray flux increased. The X-ray spectrum then continued to soften considerably in the few days after we detected the changes in the jet spectrum, transitioning from the HIMS to the SIMS between Sep 18 04:20 UT and Sep 19 08:52 UT \citep{2018MNRAS.480.4443T}.

High-cadence X-ray monitoring with the Neutron Star Interior Composition Explorer (NICER) at the same time as the onset of the compact jet quenching \citep{2018ApJ...865L..15S} showed similar results. The (1--10\,keV) source hardness continued to decrease steadily and the (3--10\,keV) X-ray variability remained approximately stable (with an rms between 15--12\%) until $\sim$1\,day after our Sep 17 radio observation ($\sim$1.5--2\,days after the IR appeared to initially fade). Again, the soft X-ray variability and hardness also changed rapidly between Sep 18 and Sep 20 \citep{2018ApJ...865L..15S}. Furthermore, a detailed analysis of the higher energy X-ray observations from AstroSAT \citep{2019MNRAS.488..720B} showed a short-lived (half a day long) decrease in the 30--80\,keV count rate on Sep 16 (although the change in count could be related to spectral shape evolution). Over the same time, there was an increase and then decrease in the X-ray quasi-periodic oscillation (QPO) frequency, from 2.1 to 3\,Hz (measured in the 3--80\,keV band). However, these hard X-ray features occurred $\sim$12\,hours before the IR emission began to fade and $\sim$1\,day before we detected \nub\ within the radio band. While these events may be related, the expected time delay between the X-ray--IR ($\sim$0.1\,seconds; see \citealt{2017NatAs...1..859G,2020arXiv200208399R}) and X-ray--radio ($\sim$30\,minutes; see \citealt{2019MNRAS.484.2987T}) changes should be much shorter.

Therefore, from the X-ray monitoring available, the clearest change to the emission from the accretion flow seems to be related to the X-ray hardness: the X-ray spectrum softened gradually leading up to the jet changes before it softened considerably in the few days after, entering the SIMS $\sim$1--2 days after our Sep 17 epoch. These changes suggest that whatever was driving the changes to the hardness may also play a role in the launching of the compact jet. Speculatively, the increase of soft X-ray photons leading up to and triggering the state change could play a role in the quenching of the compact jet emission, possibly by cooling/depleting the corona such that it was no longer able to sustain the jet. \citet{2014MNRAS.439.1381R} suggested a potential correlation between the location of \nub\ and the source hardness. While we did not observe a correlation during the jet quenching, it is plausible that the onset of these two events are connected but occur on different timescales during the jet quenching and re-ignition phases, i.e.,\ during the quenching phase the change occurs rapidly once the source reaches some critical X-ray hardness (or softness), while during the decay the change is more gradual as the corona builds up. A change in \nub\ was observed in the neutron star XRB Aquila X-1\footnote{To date, Aquila X-1 is the only neutron star system with constraints on the evolution of the jet spectral break.}, which may have also been connected to the X-ray spectral hardness of the source \citep{2018A&A...616A..23D}. If such a connection is present, it would imply a similar process may be responsible for jet key changes in the jet acceleration in both neutron star and BH systems.

\begin{figure}
\centering
\includegraphics[width=0.99\columnwidth]{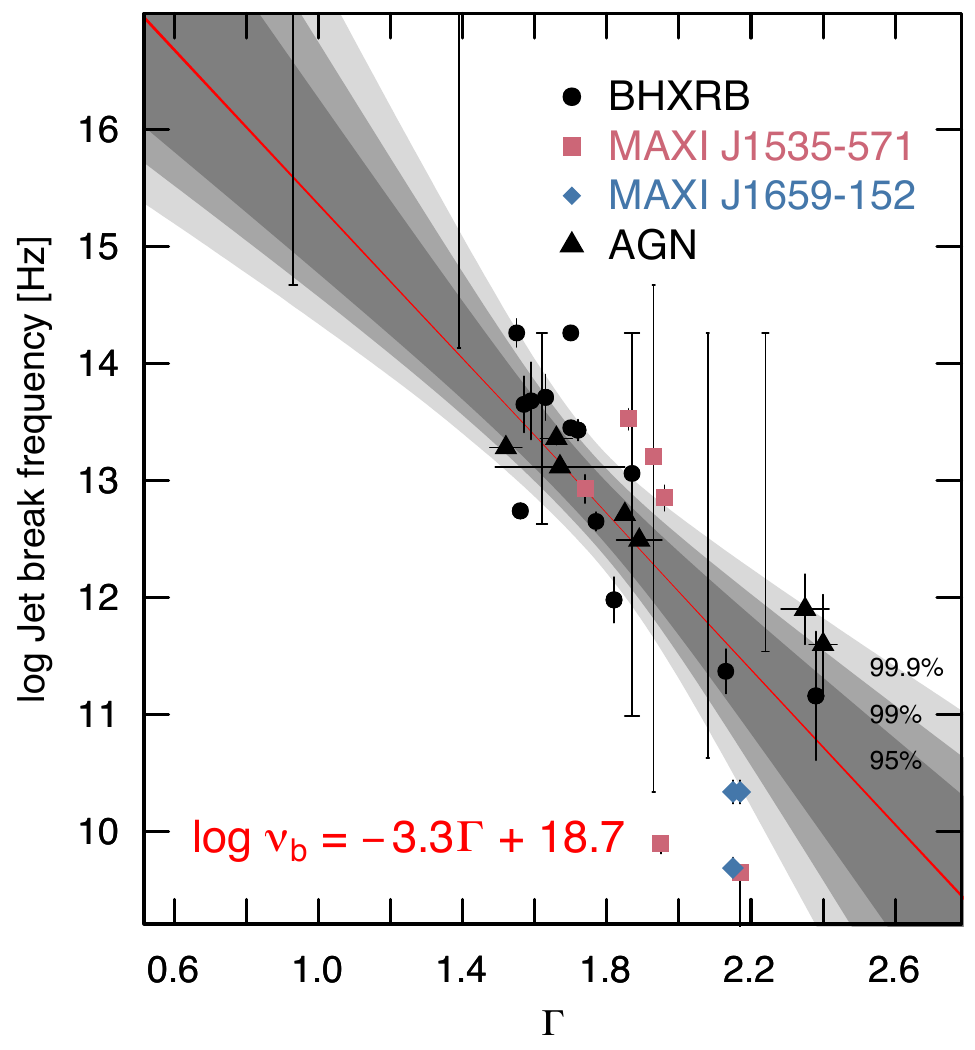}
\caption{Jet break frequency versus X-ray power law photon index (adapted from \citealt{2015ApJ...814..139K}). Measurements from this work are shown as magenta squares, MAXI~J1659$-$152 results are plotted as blue diamonds, the entire sample of jet break frequency measurements from BH XRBs are plotted as black circles, and AGN are shown as black triangles. The red line shows the median of the Monte-Carlo fit, where the relation is provided in the lower left corner. The increasingly dark shaded regions show the 95\%, 99\% and 99.9\% confidence intervals of the linear regression fits (see \citealt{2015ApJ...814..139K} for full details on the fitting). XRB data with only limits on \nub\ are shown as vertical bars. For these points, the horizontal marker caps show the 1-$\sigma$ error on $\Gamma$. Results from our multiwavelength monitoring of \source\ show that our first four \nub\ measurements broadly agree with the general population. However, as \nub\ decreased, we observed lower values of $\Gamma$ than for most systems, occupying a similar parameter space as MAXI~J1659$-$152 (blue diamonds) when \nub\ was detected within the radio band as the compact jet was beginning to fade.}
\label{fig:koljonen}
\end{figure}

An empirical correlation between \nub\ and the X-ray photon index, $\Gamma$, has been proposed for both accreting stellar-mass and supermassive BHs (\citealt{2015ApJ...814..139K}, but see \citealt{2018A&A...614L...5K} for theoretical approximations). This relation implies that the internal properties of the jet are connected to the conditions of the plasma close to the black hole. Exploring this connection for \source\ shows that $\Gamma$ and \nub\ from the first four multiwavelength epochs of \source\ (Sep 12--Sep 16) agree well with the broader population of BH XRBs and AGN (at lower values of $\Gamma$; Figure~\ref{fig:koljonen}). However, as the source softened (on Sep 17) and the radio jet began to rapidly evolve, \nub\ appeared lower than expected for that value of $\Gamma$, which continued for our observation on Sep 21, when the break was below the radio band\footnote{As mentioned previously, the Sep 17 X-ray observation was taken $\sim$10\,hours before our radio observation, so we also compared \nub\ to the X-ray photon index measured on Sep 18, and find a similar result, where \nub\ falls below the expected correlation.}. Interestingly, our result for this epoch lies in the same region as the results determined for MAXI~J1659$-$152 \citep{2013MNRAS.436.2625V} during a similar phase of the outburst (Figure~\ref{fig:koljonen}), as opposed to the majority of the other measurements which have been taken during the outburst decay phase when the jet is re-igniting. The similarities in the evolution of \nub\ for both \source\ and MAXI~J1659$-$152 suggests that the process driving the changes may be the same between these two sources. However, we note that the relation proposed by \citet{2015ApJ...814..139K} used only observations with broad X-ray coverage (much greater than 10\,keV) to ensure a well constrained X-ray photon index. Our X-ray data were only taken between 0.5--10\,keV and we find discrepancies in $\Gamma$ with some X-ray telescopes (but not others; Section~\ref{sec:x-ray}). Therefore, a caveat to this comparison is the narrow X-ray energy coverage in the \swift\ data used.

\subsection{Connection to the transient jet}

Close in time to the onset of the compact jet quenching, \source\ launched a transient jet \citep{2019ApJ...883..198R}, which produced bright optically-thin radio flares a few days after our Sep 17 radio observations \citep{2019MNRAS.488L.129C,2019ApJ...883..198R}. Due to optical depth effects, the timing of the radio flares are not indicative of the flow of material down the jet \citep[e.g.,][]{2009MNRAS.396.1370F,2012MNRAS.421..468M}. However, spatially tracking the jet knot as it propagated outwards from the system and extrapolating its motion back in time allowed the launching time to be constrained to between Sep 12 and Sep 17 (where the extent is shown as the vertical dashed lines in Figure~\ref{fig:rfaz}). As such, the timing of the rapid compact jet quenching falls within the range determined for the launching of the transient jet indicating a possible connection between the two events. 

\citet{2003ApJ...597.1023V} suggested that the transient jet may arise from the ejection of the corona (see also discussions in \citealt{2003ApJ...595.1032R}). Naively, in this scenario the transient jet and quenching of the compact jet would be directly connected, such that the compact jet emission switches off when the corona disappears and a transient jet is launched. However, during the 2011 outburst of MAXI~J1836$-$194, \nub\ was observed to shift to lower frequencies as the compact jet faded despite the system remaining in the hard/hard-intermediate X-ray spectral states with no ejection events \citep{2013ApJ...768L..35R,2014MNRAS.439.1390R,2015MNRAS.450.1745R}. Similarly, despite showing similar \nub\ behaviour, no ejecta were launched from the BH XRB MAXI~J1659$-$152 \citep{2013MNRAS.432.1319P}. Radio monitoring also showed \nub\ to move down into the radio band before shifting back to higher frequencies \citep{2013MNRAS.436.2625V}, highlighting its non-linear evolution. IR observations of GX~339$-$4 \citep{2011ApJ...740L..13G} also implied that \nub\ had shifted below and above the IR observing band rapidly. These results suggest that while the transient jet may well arise from the ejection of the corona, which may in-turn quench the compact jet, the quenching of the compact jet does not strictly require the launching of the transient jet.

Alternatively, as discussed in Section~\ref{sec:timescales}, it may be that the compact jet emission switched off as the jet acceleration region disconnected from the system. If that was occurring, we might speculate that the transient jet ejection may be the now-disconnected dissipation region propagating away from the system on the dynamical timescales of the flow. In such a scenario, for cases where the \nub\ was observed to evolve to lower and then back to higher frequencies (the acceleration regions moving away from and then back towards the BH; e.g.,\ \citealt{2011ApJ...740L..13G,2013MNRAS.436.2625V}), the acceleration region may not have completely disconnected from the flow, and variations in the flow of material into the jet base re-ignites the jet. As a source moves back out of a typical soft state (lasting a few weeks to months) where the jet emission was not detected, the particle acceleration region may have moved sufficiently far away from the system that it may recover in a different way, resulting in the gradual re-launching of the compact jet \citep[e.g.,][]{2013ApJ...779...95K,2014MNRAS.439.1390R}. However, this scenario may be difficult to reconcile with multiple jet ejections that may be observed from BH XRBs \citep[e.g.,][]{1994Natur.371...46M,1999MNRAS.304..865F,2017MNRAS.469.3141T,2019Natur.569..374M}. However, speculatively, multiple ejections could result from rapid re-launching and switching off of the compact jet, which could produce short IR flaring events that have been observed. High resolution Very Long Baseline Interferometry (VLBI) tracking ejecta as they are launched, or detailed radio monitoring (preferably sensitive to the low surface brightness temperatures of the expanding jet knots) tracking the ejecta as they propagate out \citep{2019ApJ...883..198R,2020NatAs.tmp....2B} will be crucial to understanding if these events are connected.

\subsection{Comparison between the compact jet quenching and re-launching timescales}

At the end of its outburst \source\ transitioned from the soft to hard state at an unusually low X-ray luminosity \citep{2019MNRAS.488L.129C,2019ApJ...883..198R} and no radio emission was detected. Therefore, we were unable to compare the processes of jet quenching and re-launching in this source during its major outburst. 

However, \source\ underwent a number of radio and X-ray re-brightenings following the end of its major outburst \citep{2019ApJ...878L..28P}. During these re-brightenings the source transitioned between the hard and soft states, and during the latter the compact jet was quenched. While lack of multiwavelength monitoring precludes constraints on the frequency of the jet spectral break, dense radio monitoring of the source as it completed a soft to hard transition showed the radio spectrum gradually had evolved from steep to flat/inverted over $\sim$1\,week (see \citealt{2019ApJ...878L..28P} for full details). While \nub\ was not constrained, this slow change to the radio spectrum indicated a more gradual evolution for the re-launching.

The only other system where the time-evolution of \nub\ has been measured as it moved through different observing bands was during the hard state decay of the 2011 outburst of the BH XRB MAXI~J1836$-$194. Multi-wavelength monitoring of that source indicated that \nub\ evolved from $\sim$10$^{11}$\,Hz to $\sim$10$^{14}$\,Hz ($\sim$3 orders of magnitude) over a period of $\sim$6\,weeks as the source hardened during the outburst decayed \citep{2013ApJ...768L..35R,2014MNRAS.439.1390R}.

The contrast between the timescales of the compact jet quenching and re-launching in this outburst is suggestive of a different mechanism driving these two processes. Such a difference may result in the observed difference in the radio/X-ray coupling during the hard-state rise and decay phases of an outburst \citep{2018MNRAS.481.4513I,2019ApJ...871...26K}. We highlight that with reasonable constraints from only two sources, one quenching and the other re-launching, this observed behaviour may not represent the BH XRB population. In particular, the outburst of MAXI~J1836$-$194 did not enter the soft state and the compact jet did not quench \citep{2015MNRAS.450.1745R}. Therefore, the slow evolution of \nub\ may have been related to jet recovery, as opposed to the complete relaunching of the compact jet. However, while we did not have good multiwavelength constraints, the slow $\sim$1-week evolution of the radio spectrum during soft-state re-brightenings from \source\ (after the major outburst) does suggest similar behaviour \citep{2019ApJ...878L..28P}. During these re-brightenings, \source\ did enter the soft state and the radio jet was not detected. As the re-flare faded and the source returned to a hard state the radio emission brightened, and slowly evolved from a steep to inverted radio spectrum over a $\sim$1-week timescale. 

A slow evolution of \nub\ was also implied by dense radio monitoring during the soft-to-hard state transition of the 2011 outburst from GX~339$-$4, where the radio spectrum showed a smooth and gradual evolution ($\alpha \approx -0.6$--$+$0.3) over a $\sim$2\,week period, with the IR peaking at least a few days later \citep{2013MNRAS.431L.107C}. This more gradual evolution is also supported by the $\sim$2-week IR rise-time following the soft$\rightarrow$hard state transitions in 4U~1543$-$47 and XTE~J1550$-$564 (see discussions in \citealt{2013ApJ...779...95K}). In contradiction, recent monitoring of the BH XRB 4U~1543$-$47 showed a sudden increase in IR emission during a short ($\sim$5\,day) return back to the SIMS from a soft state\footnote{Over a soft state $\rightarrow$ SIMS $\rightarrow$ soft state transition.}. Such flaring may indicate that the compact jet emission had switched back on quickly and the jet spectral break shifted rapidly back up to the IR band. However, these campaigns generally lacked comprehensive and simultaneous multiwavelength support. As such, further observations that include simultaneous (or as close to simultaneous as possible) radio, mm, and IR observations of the jet as it is switching off and re-igniting (preferably after deep quenching in a soft state) are necessary to understand how jets are launched and quenched by the accretion flow, and how/if these two processes differ.

\section{Conclusions}
With almost daily multi-frequency monitoring of \source\ over its HIMS$\rightarrow$SIMS transition, we detected the onset of rapid compact jet quenching, as the higher energy jet emission faded in less than 1\,day. The jet quenching occurred a few days before the transition to the SIMS. Over this time the jet spectral break decreased in frequency by $\sim$3 orders of magnitude, from the IR band into the radio band. While we are unable to identify any direct causal changes in the X-ray variability, the X-ray photon index and X-ray spectrum gradually softened leading up to this event, before a more rapid change in the days afterwards. Therefore, it is possible that the increase in the soft X-ray photons and decrease in the hard X-ray emission may be connected to the observed changes in the jet. We find that the rapid jet quenching began at a similar time to the launching of a transient jet ejection, possibly connecting the two events.

From a time and frequency analysis of our radio observations, we show that on 2017 Sep 17, \nub\ was within the radio band and decreased by $\sim$1.8\,GHz over $\approx$15\,mins. We argue that the time-evolution of \nub\ was 2--3 orders of magnitude faster than expected from synchrotron cooling, but is similar to dynamical timescales of material flowing down the jet. Therefore, our results suggest that the onset of compact jet quenching in \source\ was not driven by details of local particle acceleration; instead it appears as if internal jet properties changed dramatically and the particle acceleration region suddenly moved away from the BH with the jet flow.

Our results suggests that the mechanism resulting in the quenching and re-launching of compact jets from accreting BHs may arise from different processes. Although a more-rapid jet re-ignition may be possible, but this scenario has not yet been inferred. This work highlights the need for high-cadence radio, mm, IR, and X-ray monitoring of these objects during both the outburst rise and decay to understand how jets are launched and quenched from accretion flows, and whether that process is universal between different sources or at different times within the same source.

\section*{Acknowledgements}

We thank the anonymous referee for their helpful comments. We also thank Jamie Stevens and staff from the Australia Telescope National Facility (ATNF) for scheduling the ATCA radio observations, as well as the \swift\ team for the scheduling of the X-ray observations. AJT thanks Gerald Schieven for his help configuring and scheduling the ALMA observations reported in this work. TDR acknowledges support from the Netherlands Organisation for Scientific Research (NWO) Veni Fellowship, grant number 639.041.646. M. L. and S. M. are thankful for support from an NWO VICI award, grant Nr. 639.043.513. JCAM-J is the recipient of an Australian Research Council Future Fellowship (FT140101082), funded by the Australian government. AJT acknowledges support for this work through an Natural Sciences and Engineering Research Council of Canada (NSERC) Post-Graduate Doctoral Scholarship (PGSD2-490318-2016). GRS and AJT acknowledge support from an NSERC Discovery Grant (RGPIN-06569-2016). FK was supported as an Eberly Research Fellow by the Eberly College of Science at the Pennsylvania State University. KIIK was supported by the Academy of Finland project 320085. JvdE, ASP and ND are supported by a NWO Vidi grant, awarded to ND. This research has made use of \textsc{ISIS} functions (\textsc{ISIS}scripts) provided by ECAP/Remeis observatory and MIT (\url{http://www.sternwarte.uni-erlangen.de/isis/}). We thank J. E. Davis for the development of the \texttt{slxfig} module that was used to prepare some of the figures in this work.  The Australia Telescope Compact Array (ATCA) is part of the Australia Telescope National Facility which is funded by the Australian Government for operation as a National Facility managed by CSIRO. We acknowledge the Gomeroi people as the traditional owners of the ATCA Observatory site. This paper makes use of the following ALMA data: ADS/JAO.ALMA\#2016.1.00925.T. ALMA is a partnership of ESO (representing its member states), NSF (USA) and NINS (Japan), together with NRC (Canada), MOST and ASIAA (Taiwan), and KASI (Republic of Korea), in cooperation with the Republic of Chile. The Joint ALMA Observatory is operated by ESO, AUI/NRAO and NAOJ. The National Radio Astronomy Observatory is a facility of the National Science Foundation operated under cooperative agreement by Associated Universities, Inc.

\section*{Data Availability}
Data from \swift\ are publicly available from HEASARC (\url{https://heasarc.gsfc.nasa.gov/}). Raw ATCA data are provided on the Australia Telescope Online Archive (\url{https://atoa.atnf.csiro.au/query.jsp}). Raw ALMA data are available online at the ALMA Science Archive (\url{https://almascience.eso.org/asax/}). All calibrated flux densities used in this work are provided in Table~\ref{tab:data} and Table~\ref{tab:sep17_radio_data}.



\bibliographystyle{mnras}
\bibliography{bib}


\onecolumn
\appendix

\section{Quasi-simultaneous data}
All quasi-simultaneous radio, sub-mm, IR and optical data used in this work are provided in Table~\ref{tab:data}.

\begin{center}
\begin{longtable}{ccccc}
\caption {Radio, sub-mm, IR and optical data of the 2017 outburst of \source. All radio data are from \citet{2019ApJ...883..198R}, sub-mm data are new to this work, and the de-reddened IR/optical are from \citet{2018ApJ...867..114B}. The epoch relates to the 2017 date of the radio data (and is the same as Table~\ref{tab:pars}). For precise date, use MJD. Errors on the ATCA radio data are dominated by the estimated systematic uncertainties of 4\% (see Section~\ref{sec:radio}). To account for variability, errors on the ALMA sub-mm data are larger than the expected  systematic uncertainties of 5\% (see Section~\ref{sec:alma}).} 
\label{tab:data} \\

\hline 
\multicolumn{1}{c}{Epoch} & \multicolumn{1}{c}{MJD}  & \multicolumn{1}{c}{Telescope} & \multicolumn{1}{c}{Central frequency} & \multicolumn{1}{c}{$S_{\nu}$} \\

\multicolumn{1}{c}{} & \multicolumn{1}{c}{} & \multicolumn{1}{c}{} & \multicolumn{1}{c}{(Hz)} & \multicolumn{1}{c}{(mJy)} \\ \hline

\endfirsthead

\multicolumn{5}{c}%
{{\tablename\ \thetable{} -- Continued from previous page. Flux densities of \source.}} \\

\hline 

\multicolumn{1}{c}{Epoch} & \multicolumn{1}{c}{MJD}  & \multicolumn{1}{c}{Telescope} & \multicolumn{1}{c}{Central frequency} & \multicolumn{1}{c}{$S_{\nu}$} \\

\multicolumn{1}{c}{} & \multicolumn{1}{c}{} & \multicolumn{1}{c}{} & \multicolumn{1}{c}{(Hz)} & \multicolumn{1}{c}{(mJy)}  \\ \hline

\endhead
\hline
 \multicolumn{5}{c}{{Continued on next page}} \\ \hline
\endfoot

\hline
\endlastfoot

2017 Sep 12 & 58008.57 & ATCA & 17$\times$10$^9$ & 172$\pm$7 \\
       & 58008.57 & ATCA & 19$\times$10$^9$ & 171$\pm$7 \\
       & 58007.91 & ALMA & 97.5$\times$10$^9$ & 232$\pm$10 \\
       & 58007.98 & ALMA & 145.0$\times$10$^9$ & 227$\pm$15 \\
       & 58007.94 & ALMA & 236.0$\times$10$^9$ & 220$\pm$20 \\

       & 58008.09 & REM & 1.39$\times$10$^{14}$ & 54$\pm$5 \\
       & 58008.09 & REM & 1.81$\times$10$^{14}$ & 45$\pm$6 \\
       & 58008.09 & REM & 2.43$\times$10$^{14}$ & 45$\pm$9 \\
       & 58008.42 & LCO & 3.94$\times$10$^{14}$ & 64$\pm$27 \\

 \hline

2017 Sep 14 & 58010.56 & ATCA & 5.5$\times$10$^9$ & 185$\pm$8 \\ 
       & 58010.56 & ATCA & 9.0$\times$10$^9$ & 185$\pm$8 \\
       & 58010.56  & ATCA & 17.0$\times$10$^9$ & 179$\pm$7 \\
       & 58010.56  & ATCA & 19.0$\times$10$^9$ & 166$\pm$7 \\

       & 58010.09 & REM & 1.39$\times$10$^{14}$ & 80$\pm$8 \\
       & 58010.09 & REM & 1.81$\times$10$^{14}$ & 60$\pm$10 \\
       & 58010.09 & REM & 2.43$\times$10$^{14}$ & 54$\pm$10 \\
       & 58010.80 & LCO & 2.99$\times$10$^{14}$ & 57$\pm$19 \\
       & 58010.09 & REM & 3.28$\times$10$^{14}$ & 49$\pm$18 \\
       & 58010.09 & REM & 3.93$\times$10$^{14}$ & 100$\pm$45 \\

 \hline

2017 Sep 15 & 58011.56 & ATCA & 5.5$\times$10$^9$ & 166$\pm$7 \\ 
       & 58011.56 & ATCA & 9.0$\times$10$^9$ & 182$\pm$7 \\
       & 58011.56  & ATCA & 17.0$\times$10$^9$ & 179$\pm$7 \\
       & 58011.56  & ATCA & 19.0$\times$10$^9$ & 175$\pm$7 \\

       & 58010.97 & VISIR & 6.19$\times$10$^{13}$ & 96$\pm$8 \\

       & 58011.10 & REM & 1.39$\times$10$^{14}$ & 65$\pm$7 \\
       & 58011.10 & REM & 1.81$\times$10$^{14}$ & 53$\pm$7 \\
       & 58011.09 & REM & 2.43$\times$10$^{14}$ & 59$\pm$11 \\

 \hline

2017 Sep 16 & 58012.55  & ATCA & 5.5$\times$10$^9$ & 164$\pm$7 \\ 
       & 58012.55  & ATCA & 9.0$\times$10$^9$ & 178$\pm$7 \\
       & 58012.53  & ATCA & 17.0$\times$10$^9$ & 184$\pm$7 \\
       & 58012.53  & ATCA & 19.0$\times$10$^9$ & 184$\pm$7 \\
 
 .
       & 58011.97 & VISIR & 2.47$\times$10$^{13}$ & 157$\pm$5 \\
       & 58013.03 & VISIR & 3.44$\times$10$^{13}$ & 141$\pm$12 \\ 
 
       & 58012.10 & REM & 1.39$\times$10$^{14}$ & 56$\pm$6 \\
       & 58012.10 & REM & 1.81$\times$10$^{14}$ & 52$\pm$7 \\
       & 58012.11 & REM & 2.43$\times$10$^{14}$ & 54$\pm$10 \\

 \hline
      
2017 Sep 17 & 58013.55  & ATCA & 5.5$\times$10$^9$ & 135$\pm$6 \\ 
       & 58013.55  & ATCA & 9.0$\times$10$^9$ & 142$\pm$6 \\
       & 58013.55   & ATCA & 17.0$\times$10$^9$ & 123$\pm$5 \\
       & 58013.55   & ATCA & 19.0$\times$10$^9$ & 118$\pm$5 \\

       & 58013.9   & VISIR & 6.19$\times$10$^{13}$ & 45$\pm$5 \\

 \hline

2017 Sep 21 & 58017.46 & ATCA & 5.5$\times$10$^9$ & 151$\pm$6 \\ 
       & 58017.46 & ATCA & 9.0$\times$10$^9$ & 121$\pm$5 \\
       & 58017.46 & ATCA & 17.0$\times$10$^9$ & 92$\pm$4 \\
       & 58017.46 & ATCA & 19.0$\times$10$^9$ & 86$\pm$3 \\
       & 58017.98 & ALMA & 97.5$\times$10$^9$ & 52$\pm$10 \\
       & 58018.05 & ALMA & 145.0$\times$10$^9$ & 30$\pm$3 \\
       
       & 58018.99 & VISIR & 3.44$\times$10$^{13}$ & 20$\pm$4 \\

       & 58017.03 & REM & 1.39$\times$10$^{14}$ & 16$\pm$3 \\
       
       & 58017.44 & LCO & 3.12$\times$10$^{14}$ & 30$\pm$9 \\
       & 58017.41 & LCO & 3.93$\times$10$^{14}$ & 52$\pm$24 \\
       & 58017.44 & LCO & 3.93$\times$10$^{14}$ & 48$\pm$21 \\

\end{longtable}

\end{center}

\section{Intra-observational radio variability for 2017 Sep 17}
\label{App:Sep17}
Radio intra-observational variability of \source\ from our ATCA observation on 2017 Sep 17, from Section~\ref{sec:variability}.

\begin{center}
\begin{longtable}{ccc}
\caption {Two-minute time-interval radio data from \source\ on 2017 Sep 17 (as shown in Figure~\ref{fig:inradio}). The MJD denotes the mid-point of each two minute interval. Bandwidths of the higher frequency ($>$15\,GHz) radio data are 1\,GHz, centered around the tabulated frequency, while the lower frequency ($<$15\,GHz) radio data have a bandwidths of 512\,MHz. Flux densities have had a 0.5\% systematic uncertainty applied (see Section~\ref{sec:radio}).} 
\label{tab:sep17_radio_data} \\

\hline 
\multicolumn{1}{c}{MJD}  &  \multicolumn{1}{c}{Central frequency} & \multicolumn{1}{c}{$S_{\nu}$}  \\

 \multicolumn{1}{c}{} & \multicolumn{1}{c}{(GHz)} & \multicolumn{1}{c}{(mJy)} \\ \hline

\endfirsthead

\multicolumn{3}{c}%
{{\tablename\ \thetable{} -- Continued from previous page. Intra-observational radio flux densities taken on Sep 17.}} \\

\hline 

 \multicolumn{1}{c}{MJD}  &  \multicolumn{1}{c}{Central frequency} & \multicolumn{1}{c}{$S_{\nu}$}  \\

\multicolumn{1}{c}{} & \multicolumn{1}{c}{(GHz)} & \multicolumn{1}{c}{(mJy)} \\ \hline

\endhead
\hline
\multicolumn{3}{c}{{Continued on next page}} \\ \hline
\endfoot

\hline
\endlastfoot

58013.53900 & 16.5 & 125.35 $\pm$ 0.63 \\
 & 17.5 & 123.83 $\pm$ 0.62 \\
 & 18.5 & 121.68 $\pm$ 0.61 \\
 & 19.5 & 118.54 $\pm$ 0.60 \\

\hline

58013.54038 & 16.5 & 125.67 $\pm$ 0.63 \\
 & 17.5 & 123.70 $\pm$ 0.62 \\
 & 18.5 & 122.24 $\pm$ 0.61 \\
 & 19.5 & 119.52 $\pm$ 0.60 \\
 
 \hline
 
 58013.54177 & 16.5 & 125.07 $\pm$ 0.63 \\
 & 17.5 & 123.32$\pm$ 0.62 \\
 & 18.5 & 122.02 $\pm$ 0.61 \\
 & 19.5 & 119.38 $\pm$ 0.60 \\
 \hline
58013.54316 & 16.5 & 124.15 $\pm$ 0.62 \\
 & 17.5 & 123.37 $\pm$ 0.62 \\
 & 18.5 & 121.84 $\pm$ 0.61 \\
 & 19.5 & 119.02 $\pm$ 0.60 \\
 \hline

58013.54941 &  4.75 &  130.08 $\pm$ 0.65 \\
       &  5.25 &  132.31 $\pm$  0.66 \\
       &  5.75 &  134.19 $\pm$  0.67 \\
       &  6.25 &  136.17 $\pm$ 0.68 \\
       &  8.25 &  143.40 $\pm$ 0.72 \\
       &  8.75 &  143.39 $\pm$  0.72 \\
       &  9.25 &  142.09 $\pm$  0.71 \\
       &  9.75 &  139.58 $\pm$  0.70 \\

 \hline
 
58013.55080 &  4.75 &  130.01 $\pm$  0.65 \\
       &  5.25 &  133.13 $\pm$  0.67 \\
       &  5.75 &  134.91 $\pm$ 0.67 \\
       &  6.25 &  136.87 $\pm$ 0.68 \\
       &  8.25 &  144.60 $\pm$ 0.72 \\
       &  8.75 &  144.34 $\pm$  0.72 \\
       &  9.25 &  142.42 $\pm$  0.71 \\
       &  9.75 &  140.40 $\pm$  0.70 \\

   \hline

58013.55219 &  4.75 &  131.00 $\pm$ 0.65 \\
       &  5.25 &  133.39 $\pm$ 0.67 \\
       &  5.75 &  135.87 $\pm$ 0.68 \\
       &  6.25 &  138.03 $\pm$ 0.69 \\
       &  8.25 &  144.89 $\pm$ 0.72 \\
       &  8.75 &  143.44 $\pm$ 0.72 \\
       &  9.25 &  142.60 $\pm$ 0.71 \\
       &  9.75 &  140.38 $\pm$ 0.70 \\

\hline
58013.55358 &  4.75 &  129.91 $\pm$ 0.65 \\
       &  5.25 &  134.55 $\pm$ 0.67 \\
       &  5.75 &  136.77 $\pm$ 0.68 \\
       &  6.25 &  139.12 $\pm$ 0.70 \\
       &  8.25 &  146.32 $\pm$ 0.73 \\
       &  8.75 &  143.42 $\pm$ 0.72 \\
       &  9.25 &  140.89 $\pm$ 0.70 \\
       &  9.75 &  140.04 $\pm$ 0.70 \\

    \hline

58013.55497  &  4.75 &  130.72 $\pm$ 0.65 \\
        &  5.25 &  134.77 $\pm$ 0.67 \\
        &  5.75 &  137.17 $\pm$ 0.69 \\
        &  6.25 &  139.87 $\pm$ 0.70 \\
        &  8.25 &  144.11 $\pm$ 0.72 \\
        &  8.75 &  142.28 $\pm$ 0.71 \\
        &  9.25 &  139.73 $\pm$ 0.70 \\
        &  9.75 &  138.83 $\pm$ 0.69 \\

   \hline

58013.55635 &  4.75 &  129.32 $\pm$ 0.65 \\
       &  5.25 &  133.11 $\pm$ 0.67 \\
       &  5.75 &  136.89 $\pm$ 0.68 \\
       &  6.25 &  139.16 $\pm$ 0.70 \\
       &  8.25 &  142.33 $\pm$ 0.71 \\
       &  8.75 &  140.07 $\pm$ 0.70 \\
       &  9.25 &  138.92 $\pm$ 0.69 \\
       &  9.75 &  136.49 $\pm$ 0.68 \\

    \hline

58013.55751 &  4.75 &  131.10 $\pm$ 0.66 \\
       &  5.25 &  133.96 $\pm$ 0.67 \\
       &  5.75 &  135.00 $\pm$ 0.68 \\
       &  6.25 &  136.10 $\pm$ 0.68 \\
       &  8.25 &  133.36 $\pm$ 0.67 \\
       &  8.75 &  132.09 $\pm$ 0.66 \\
       &  9.25 &  131.11 $\pm$ 0.66 \\
       &  9.75 &  129.58 $\pm$ 0.65 \\

\hline

58013.56399 & 16.5 & 117.24 $\pm$0.59 \\
 & 17.5 & 115.24 $\pm$ 0.58 \\
 & 18.5 & 112.99 $\pm$ 0.56 \\
 & 19.5 & 109.40 $\pm$ 0.55 \\
 \hline
58013.56539 & 16.5 & 116.90 $\pm$ 0.58 \\
 & 17.5 & 115.09 $\pm$ 0.58 \\
 & 18.5 & 112.45 $\pm$ 0.56 \\
 & 19.5 & 108.84 $\pm$ 0.54 \\
 \hline
58013.56643 & 16.5 & 116.04 $\pm$0.58 \\
 & 17.5 & 113.99 $\pm$ 0.57 \\
 & 18.5 & 111.53 $\pm$ 0.56 \\
 & 19.5 & 108.67 $\pm$ 0.54 \\

\end{longtable}

\end{center}


\section{Radius and magnetic field of the first acceleration region}
\label{App:chaty}
Derivation of the magnetic field and radius of the first acceleration region as outlined by \citet{2011A&A...529A...3C}.  From \citet{1979rpa..book.....R} and \citet{2011hea..book.....L}, for a synchrotron-emitting source with a power law distribution of electrons with an energy spectrum $N(E) \, dE = \kappa E^{-p} \, dE$, where $p$ is the spectral index of the particle energies, the magnetic field of the first acceleration region, $B_{\rm F}$, is related to the frequency and flux density of the jet spectral break (\nub\ and $S_{\nu,{\rm b}}$, respectively), such that:
\begin{equation}
B_{\rm F} = \left( \frac{2 \mu_0^2 c^4 A h Y^2}{T^{3}}\right)^{2/(2p+13)} (D \xi)^{-4/(2p+13)} S_{\nu,{\rm b}}^{-2/(2p+13)} \sin a^{-(2p+5)/(2p+13)} \nu_{\rm b}, 
\label{eq:B_F_full}
\end{equation}
where 
\begin{equation}
A=\frac{\sqrt{3}e^3}{8 \pi^2 \epsilon_0 c m_e (p+1)} \left( \frac{m_e^3 c^4}{3 e } \right)^{-(p-1)/2} \Gamma_{\rm f} \left( \frac{p}{4}+ \frac{19}{12} \right) \Gamma_{\rm f} \left( \frac{p}{4}-\frac{1}{12} \right),
\label{eq:A}
\end{equation}
\begin{equation}
Y =  \frac{\gamma_{\rm max}^{(2-p)}-\gamma_{\rm min}^{(2-p)}}{(2-p)},
\label{eq:Y}
\end{equation}
and
\begin{equation}
T = \frac{\sqrt{3} e^3 c}{32 \pi^2 \epsilon_0 m_e} \left( \frac{3e}{2 \pi m_e^3 c^4} \right)^{p/2} \Gamma_f \left( \frac{3p+22}{12} \right) \Gamma_f \left( \frac{3p+2}{12} \right).
\label{eq:T}
\end{equation}
Here, $\mu_0$ is the permeability of free space, $c$ is the speed of light, if we assume that the synchrotron emitting accelerating region is a homogeneous cylinder of radius $R_{\rm F}$ and height $H_{\rm F}$, factor $h$ relates the two such that $H_{\rm F}=h R_{\rm F}$ where we assume $h$=1, $D$ is the distance to the source. $\xi = 1$ assumes that the energy of the non-thermal electrons equals the magnetic energy density, and $a$ is the pitch angle of the electrons where we average over an isotropic distribution of pitch angles. $e$ and $m_e$ are the charge and mass of an electron, respectively. $B$ is the magnetic field strength, $\epsilon_0$ is the permittivity of free space, $\Gamma_{\rm f}$ is the gamma function, and finally, $\gamma_{\rm min}$ and $\gamma_{\rm max}$ are the minimum and maximum Lorentz factors of the electrons, respectively, such that $\gamma_{\rm max} \gg \gamma_{\rm min} = 1$.

The cross-sectional area of the first acceleration region, $R_{\rm F}$, is described as: 
\begin{multline}
R_{\rm F} = 2 \mu_0 c^2 \left( 2 \mu_0^2 c^4 m_e^2 A h\right)^{-(p+6)/(2p+13)} T^{(p+5)/(2p+13)} (\xi Y)^{-1/(2p+13)} \sin a^{2/(2p+13)} D^{(2p+12)/(2p+13)} S_{\nu,{\rm b}}^{(p+6)/(2p+13)} \nu_{\rm b}^{-1}.
\label{eq:R_F_full}
\end{multline}


\bsp	
\label{lastpage}
\end{document}